\documentclass[useAMS,usenatbib,usegraphicx]{mn2e}

\usepackage{epsfig,amsmath,rotating}

\newcommand{\bq}{\begin{equation}}
\newcommand{\eq}{\end{equation}}
\newcommand{\bqn}{\begin{eqnarray}}
\newcommand{\eqn}{\end{eqnarray}}
\newcommand{\dd}{\mbox{\rm d}}
\newcommand{\msun}{\rm{M}_\mathrm{\rm \sun}}

\title[Milky Way disc model: Part 1]{\sc Towards a fully consistent Milky Way 
disc model: Part 1 The local model based on kinematic and photometric data.}

\author[A. Just \& H. Jahrei\ss]{A. Just$^1$, H. Jahrei\ss$^1$.\\
  $^{1}$ Astronomisches Rechen-Institut, Zentrum f\"{u}r Astronomie der 
  Universit\"{a}t Heidelberg (ZAH), \\ M\"{o}nchhofstra\ss{}e 12-14, 69120 
  Heidelberg, Germany}
\begin{document}

\date{Accepted ...}

\pagerange{\pageref{firstpage}--\pageref{lastpage}} \pubyear{2009} 


\maketitle

\label{firstpage} 


\begin{abstract}
We present a fully consistent evolutionary disc model  
of the solar cylinder. The model is based on a sequence of stellar 
sub-populations described by the star formation history (SFR) and
the dynamical heating law (given by the age-velocity dispersion 
relation AVR). The stellar sub-populations are
in dynamical equilibrium and the gravitational potential is calculated
self-consistently including the influence of the dark matter 
halo and the gas component.
The combination of kinematic data from Hipparcos and the finite lifetimes of
main sequence (MS) stars enables us to determine the detailed vertical
disc structure independent of individual stellar ages and only
weakly dependent on the IMF. 
The disc parameters are determined by applying a sophisticated best fit algorithm
to the MS star velocity distribution functions in magnitude bins.
We find that the AVR is well constrained by the local kinematics, whereas  
for the SFR the allowed range is larger.
The model is consistent with the
local kinematics of main sequence stars and fulfils the known constraints on scale
heights, surface densities and mass ratios. 
A simple chemical enrichment model is included in order to fit the local metallicity 
distribution of G dwarfs. 
In our favoured model A the power law index of the AVR is 0.375 with a minimum and
maximum velocity dispersion of 5.1\,km/s and 25.0\,km/s, respectively. The SFR shows
a maximum 10 \,Gyr ago and declines by a factor of four to the present day value of
1.5\,$\msun/\mathrm{pc}^2/\mathrm{Gyr}$.
A best fit of the IMF leads to power-law indices of 
-1.46 below and -4.16 above 1.72$\msun$ avoiding a kink at 1$\msun$.
An isothermal thick disc component with local density of $\sim 6\%$ of the stellar 
density is included. A thick disc containing more than 10\% of local stellar
mass is inconsistent with the local kinematics of K and M dwarfs.
Neglecting the thick disc component results in slight variations of the
thin disc properties, but has a negligible influence on the AVR and the
normalised SFR.
The model allows detailed predictions
of the density, age, metallicity and velocity distribution functions of MS
stars as a function of height above the mid-plane.
The complexity of the model does not
allow to rule out other star formation scenarios using the local data alone.
The incorporation of multi-band star count and kinematic data of larger samples
in the near future will improve the
determination of the disc structure and evolution significantly.  
\end{abstract}




\section{Introduction} \label{introd}

There is no 'concordance' Milky Way model available so far that
describes the structure, kinematics, chemistry and evolution of the disc in
great detail. The classical stellar density model of 
\citet{bah80a,bah80b,bah84c} composed by
a spheroid and an exponential disc with magnitude-dependent exponential scale
heights is still widely used. \citet{bah84a,bah84b} introduced a finite
set of isothermal disc components and solved the Poisson and the Jeans equation
for a dynamical equilibrium model of the vertical disc structure. At present the
most sophisticated Milky Way model based on star counts is the so-called 
'Besan\c{c}on model' of the Galaxy developed and presented in a series of 
articles. In \citet{rob03} a general description and the present
status of the model is given. In this model the thin disc is
composed by a sequence of stellar sub-populations with increasing age and
velocity dispersion.
The Besan\c{c}on model has still some serious drawbacks in the construction
of the thin disc concerning the density profiles of the components, 
the star formation history (SFR) and
the initial mass function (IMF) that will be discussed below.

The main input functions to be determined in an evolutionary disc model are the
SFR and the dynamical heating described by the 
age-velocity dispersion relation (AVR).
The $\mathrm{SFR}(t)$ of the Milky Way disc is still not very well
determined. The main reason for that is the lack of good age estimators with
corresponding unbiased stellar samples. Especially samples selected by colour
cuts or by a magnitude limit are biased with respect to the age distribution,
because there is an age-metallicity relation due to the chemical enrichment of 
the disc. Therefore not even the famous Geneva-Copenhagen sample of 14,000 F and G stars
\citep{nor04,hol09} (hereafter GCS1,GCS2) with well determined stellar 
properties 
and individual age estimates can be used to derive the star formation
history directly by star counts. Systematic biases introduced by the method of
GCS1 concerning age and stellar parameter determinations are discussed in
\citet{pon05} and \citet{hay06}. 
The Besan\c{c}on
model of the Galaxy was developed and presented in a series of articles.
In \citet{hay97a,hay97b} the SFR of the thin disc is determined 
to be approximately constant using a series of isothermal components and
applying an approximate method to achieve dynamical equilibrium. In further
applications they used a constant SFR combined with a steep IMF in the sensitive
mass regime of 1-3\,$\msun$ \citep{rob03}. 
As a result, the mean age of the disc population, the scale heights and the
surface densities of the stellar sub-populations are relatively small.
\citet{roc00} used chromospheric age determinations of late-type dwarfs. 
They apply stellar evolution and scale height corrections. The
main result is the determination of enhanced star formation episodes over the
lifetime of the disc. They exclude a constant SFR from chemical evolution
models. In \citet{fue04} the star formation history
for open star clusters was determined. 
They found at least five episodes of enhanced star formation in the last 2\,Gyr.
Since star clusters contain only a small
percentage of all disc stars, an extrapolation to the total (smoothed) SFR is
not possible. 

In recent years the Hipparcos stars with precise parallaxes and proper motions
were used to determine the SFR with different methods.
\citet{her00} determined the SFR over
the last 3\,Gyr using isochrone ages. They found a series of star formation episodes on top of  an
underlying smooth SFR. This result is complementary to our model, which gives
the slow evolution of the smoothed SFR. 
\citet{ver02} applied an inverse method to derive from the
colour magnitude diagram the star formation history and age-metallicity
relation. They used prescribed AVR
and gravitational potential and fixed the IMF to be a power law over the whole
stellar mass range. This ansatz results in a steep
IMF and a rather young thin disc. Additionally they found strong
maxima in the SFR at ages of $10^7$ and $2\times10^9$\,yr.
In \citet{bin00} and in \citet{cig06} the scale
height variation of main sequence stars was not taken into account. Therefore
these models derive the local age distribution of stellar sub-populations with
lifetimes larger than the age of the disc instead of the SFR. 
\citet{bin00}
determined the age of the solar neighbourhood to be $11.2\pm 0.75$\,Gyr, which
includes the contribution of the thick disc.
Both papers
found an approximately constant local age distribution, which corresponds to a
decreasing SFR according to the increasing scale height with increasing age of
the stellar sub-populations.
In a recent paper \citet{aum09} derived the SFR using the updated Hipparcos data
\citep{vle07} and the kinematics of the GCS sample. They calculated in detail
the number of main sequence (MS) stars as function of B-V colour based on a
KTG93-like IMF and scale height corrections due to dynamical heating. Their
favoured disc model has an exponentially decaying SFR with a decay timescale of
8.5\,Gyr and an age of 12.5\,Gyr. In the model the gravitational potential is
fixed and the SFR depends strongly on the properties of the IMF by construction.

The presented paper (Paper I) starts a new attempt to develop a fully consistent
disc model which is able to predict the properties of the Milky Way disc in great
detail concerning density distributions, age distributions, kinematics and chemistry
of all types of stars.
The aim of this sequence of papers is to construct a smooth physically
consistent disc model that
can be extended to a global Galaxy model and allows detailed predictions of
number densities and kinematics of stars of different types. 
The new model can be of some help for the construction of a
'concordance' model of the Milky Way. It can be used for the preparation of
the future astrometric space mission Gaia that will measure with high
precision positions, distances, proper motions and stellar parameters
(temperature, surface gravity and chemical composition) of one
billion stars \citep{per01,bai05}.

We start with a local disc model of the solar cylinder, which is based on the
 SFR, the AVR and a chemical evolution function described by the 
 age-metallicity-relation (AMR).
The vertical density profiles and the corresponding scale height determinations
of thin and thick disc stars are calculated in a self-consistent 
gravitational potential including the DM halo and the gas component. 
The model parameters are determined by a best fit procedure
of the velocity distribution functions $f_\mathrm{i}(|W|)$ of solar neighbourhood 
stars from B--K dwarfs. 
We use Hipparcos stars and at the faint end the Catalogue of Nearby Stars
(CNS4), select the MS stars and divide these into a series of
volume complete sub-samples in magnitude bins. The derived
$f_\mathrm{i}(|W|)$ of each sub-sample are  compared simultaneously 
to the model predictions.
Details of the fitting procedure
and the significance with respect to parameter variations are discussed
in a future paper \citep{jus09} (Paper II).

Since the velocity
distribution function of MS samples is the average over the lifetime
of these stars properly weighted by the SFR and the dilution due to the
increasing scale height, the time resolution is strongly limited. Therefore we
use only smooth input functions for the SFR, AVR and metal enrichment in the model.
The resulting SFR, AVR
and chemical evolution describe the long-term smoothed disc evolution.
Similar disc models by fitting the vertical luminosity
and colour profiles of edge-on galaxies instead of the kinematics have already 
been used successfully \citep{jus96,jus06}. 

The main advantages of this method is that it does not depend on individual
stellar age determinations and that it is essentially independent of the shape of
the IMF. A weakness is that we cannot confine the SFR strongly based on local data
only. The application of the model to star counts of remote stellar populations will
be of great help in this respect.
In subsequent work the local
model will be continuously extended and compared to large data samples taken from 
the Sloan Digital Sky Survey SDSS/SEGUE \citep{aba09}
and the Radial Velocity Experiment RAVE \citep{zwi08} to further constrain the
possible parameter range of the disc model.

A major restriction of a local model is that radial mixing
of stars in the disc cannot be included easily. 
There are two main mixing processes
discussed in the literature. The first one is directly connected to the
dynamical heating process responsible for the AVR. The gravitational scattering
process leads to a diffusion of orbits in velocity space (AVR) and in
position (radial, tangential, vertical mixing). 
\citet{wie96} discussed the radial diffusion of stellar orbits and the
consequences for the AMR in some
detail. The radial probability distribution function of the birth-places
depends mainly on the age of the stellar sub-population and only weakly on the physical 
scattering process. For a 5\,Gyr old population the typical radial width of the
initial distribution is $\pm 2$\,kpc.
The second mechanism is resonant scattering of circular
orbits by spiral arms \citep{sel02,ros08,sch09}. This mechanism changes the
radial position of stars quickly, but the eccentricity of the orbits
remain very small.
The effect of resonant scattering may account for up to half the stars
in the solar neighborhood  \citep{ros08}, which shows the possible scale of the 
errors which may result if migration is ignored.
Unfortunately the efficiency and the properties of resonant scattering in the
Milky Way disc are still poorly known. In a future global disc model
radial mixing may be included in a parametrized form. The local model presented
in this paper is primarily a description of the 'status quo' of the local Milky
Way disc quantified by the local age distribution (SFR), the kinematics 
(AVR) and the chemical composition (AMR). 
Therefore the relations between these function are not altered by radial mixing.
But the physical interpretation of the SFR as a 
'local star formation history' and the AVR as a 'local dynamical heating process'
are weakened by radial mixing and require corrections. For the chemical
evolution model the consequences are more severe. The tight connection of the 
enrichment, the SFR and the gas infall will be broken. 
The chemical enrichment of the gas decouples partly from the (apparent)
enrichment of the stellar population. A more general local AMR including an
intrinsic scatter would be the consequence as derived by \citet{sch09}.

In section \ref{model} the physics of the disc model is presented,
section \ref{observ} describes the observational data, 
in section \ref{disc} the fitting procedure and the properties of the best fit
models are given, section \ref{summary} collects the main results and discussion.

\section{Disc Model} \label{model}

In this section we describe the construction of the disc model and the
determination of the properties of the stellar sub-samples. 
We use a thin, self-gravitating disc composed of a sequence of stellar
sub-populations according to the SFR and the AVR as input functions. 
Additionally the gravitational force of the gas component and the DM halo 
are included. The bulge and the stellar halo can be neglected in the solar
neighbourhood.
 The total gravitational potential and 
the density profile of each age-bin are determined self-consistently assuming
isothermal distribution functions with velocity dispersion according to the
AVR.
We take into account finite stellar lifetimes and mass loss due to stellar
evolution. The stellar remnants stay in their sub-population, the expelled gas
(stellar winds, PNs, and SNs) is mixed implicitly into the gas component.
Since the lifetimes and mass loss depend on metallicity, the metal
enrichment with time is included. A standard  IMF is used for the determination of
the stellar mass fraction of the sub-populations with age.
 The velocity distribution functions $f_\mathrm{ms}(|W|)$ for main sequence stars 
are calculated by a superposition of the Gaussians weighted by the local density
contribution up to the lifetime of the stars.

\subsection{Self-gravitating disc \label{grav}}

The backbone of the disc is a self-gravitating vertical disc profile 
in the thin disc approximation including 
the gas component and the DM halo. In this approximation the
Poisson-Equation is one-dimensional
and in the case of a purely self-gravitating thin disc
(i.e. no external potential) the Poisson equation can be
integrated leading to
\bq
\left(\frac{\dd\Phi}{\dd z}\right)^2
        =8\pi G \int_0^{\Phi}\rho(\Phi')\dd\Phi' \quad.\label{eqkz}
\eq
Therefore we model all gravitational components by a thin disc approximation.
We include in the total
potential $\Phi$ the contributions of the stellar component $\Phi_\mathrm{s}$, 
the gas component $\Phi_\mathrm{g}$, 
and the DM halo $\Phi_\mathrm{h}$
\bq
\Phi(z)=\Phi_\mathrm{s}(z)+\Phi_\mathrm{g}(z)+\Phi_\mathrm{h}(z)\quad.\label{eqpot}
\eq
In order to obtain the force of a
spherical halo correctly in the thin disc approximation,
 we use a special approximation (see subsection \ref{halo}).
Since we construct the disc in dynamical 
equilibrium, the density of the sub-components are given as a function 
of the total potential $\rho_\mathrm{j}(\Phi)$.
The vertical distribution is given by the implicit function $z(\Phi)$ 
via direct integration
\bq
z(\Phi)=\int_0^{\Phi}\dd\Phi'
        \left[8\pi G \int_0^{\Phi'}\rho(\Phi'')\dd\Phi''\right]^{-1/2}
         \quad.\label{eqz}
\eq
For the fitting procedure it is essential to separate the normalised model and
the scaling factors. 
Since we use the SFR and the AVR as input functions, it is convenient to
normalise the model to the total surface density $\Sigma_\mathrm{tot}$ 
and to the
maximum velocity dispersion $\sigma_\mathrm{e}$ of the oldest thin disc
sub-population. As a consequence the gravitational potential in 
equations \ref{eqkz} and \ref{eqz} is normalised to
$\sigma_\mathrm{e}^2$ and a natural scale length is
\bq
z_\mathrm{e}=\frac{\sigma_\mathrm{e}^2}
			{2\pi G \Sigma_\mathrm{tot}}, \label{eqze}
\eq
which corresponds to the exponential scale height of an isothermal 
component with velocity dispersion $\sigma_\mathrm{e}$ 
above a gravitating sheet with total surface density $\Sigma_\mathrm{tot}$. 

The normalised disc model describes the intrinsic structure of the disc.
 The relative thin disc, thick disc, gaseous, and DM fractions of 
  the surface density (up to $|z|=z_\mathrm{max}$) are given  by the input parameters
$Q_\mathrm{s},Q_\mathrm{t},Q_\mathrm{g},Q_\mathrm{h}$ that have to be iterated to find the
best model.
The value of  $z_\mathrm{max}$ is determined by the prescribed maximum of the
normalised potential $\Phi(z_\mathrm{max})/\sigma_\mathrm{e}^2$.
The disc model has two free global scaling parameters to convert the normalised model
to physical quantities. One can choose two parameters out of four types: a local
density, a scale height or thickness, a surface density, or a velocity
dispersion (see section \ref{disc}).

\subsection{Stellar disc \label{stars}}

The stellar component is composed of a sequence of isothermal sub-populations
characterised by the IMF, the chemical enrichment $\mathrm{[Fe/H]}(t)$, 
the star formation history $\mathrm{SFR}(t)$, and the dynamical evolution described by
the vertical velocity dispersion $\sigma_\mathrm{W}(\tau)$. 
Throughout the paper we use $t$ for the time with 
present time $t_\mathrm{p}=12$\,Gyr and
$\tau=t_\mathrm{p}-t$ for the age of the
sub-populations. 
We include mass loss due to stellar evolution and retain the
stellar-dynamical mass fraction $g(\tau)$ (stars + remnants) only. The mass lost
by stellar winds, supernovae and planetary nebulae is mixed implicitly to the
gas component. 

With the Jeans equation the vertical
distribution of each isothermal sub-population is given by
\bq
\rho_\mathrm{s,j}(z)=\rho_\mathrm{s0,j}
	\exp\left( \frac{-\Phi(z)}{\sigma^2_\mathrm{W,j}}\right)
\quad,
\eq
where $\rho_\mathrm{s,j}$ is actually a 'density rate', the density per age bin,
and $\sigma_\mathrm{W,j}$ the velocity dispersion at age $\tau_\mathrm{j}$. The
connection to the SFR is given by the integral over $z$
\bq
g(\tau_\mathrm{j})\mathrm{SFR}(t_\mathrm{j})=
	\int_\mathrm{-\infty}^{\infty}\rho_\mathrm{s,j}(z)\dd z\quad.
\eq
with time $t_\mathrm{j}=t_\mathrm{p}-\tau_\mathrm{j}$.
The (half-)thickness $h_\mathrm{age}(\tau_\mathrm{j})$ of the sub-populations is 
defined by the mid-plane density $\rho_\mathrm{s0,j}$ through
\bq
\rho_\mathrm{s0,j}=\frac{g(\tau_\mathrm{j})\mathrm{SFR}(t_\mathrm{j})}
			{2h_\mathrm{age}(\tau_\mathrm{j})}\quad .
\eq
and can be calculated by
\bq
h_\mathrm{age}(\tau_\mathrm{j})=\int_0^{\infty}
	\frac{\rho_\mathrm{s,j}(z)}{\rho_\mathrm{s0,j}}\dd z=\int_0^{\infty}
	\exp\left( \frac{-\Phi(z)}{\sigma^2_\mathrm{W,j}}\right)\dd z \quad.
\eq
The total stellar density $\rho_\mathrm{s}(z)$ and 
velocity dispersion $\sigma_\mathrm{s}(z)$ are determined by
\bqn
\rho_\mathrm{s}(z)&=&\int_0^{t_\mathrm{p}}\rho_\mathrm{s,j}(z)\dd t \\
\sigma^2_\mathrm{s}(z)&=&\frac{1}{\rho_\mathrm{s}(z)}\int_0^{t_\mathrm{p}}
	\sigma^2_\mathrm{W}(\tau_\mathrm{j})\rho_\mathrm{s,j}(z)\dd t \quad.
\eqn
The stellar surface density $\Sigma_\mathrm{s}$,
which includes luminous stars and stellar remnants,
 is related to the
central density $\rho_\mathrm{s,0}$ by the (half-)thickness $h_\mathrm{eff}$
and can also be
converted to the integrated star formation $S_0$ using the effective
stellar-dynamical mass fraction $g_\mathrm{eff}$
\bqn
\Sigma_\mathrm{s}&=&\int\rho_\mathrm{s}(z)\dd z
	=2h_\mathrm{eff}\rho_\mathrm{s0} =g_\mathrm{eff}S_0
	\label{eqSigmas}\\
S_0&=&\int_0^{t_\mathrm{p}} \mathrm{SFR(t)}\, \dd t
	=\langle SFR \rangle t_\mathrm{p}\label{eqS0}\\ 
g_\mathrm{eff}&=&\frac{1}{S_0}\int_0^{t_\mathrm{p}} g(\tau)\mathrm{SFR}(t)\dd t
\label{eqgeff}
\eqn

The general shape of $\rho_\mathrm{s}(z)$ can be characterised by two shape
parameters which are $h_\mathrm{eff}$ and the exponential scale height 
$z_\mathrm{s}$ well above the plane. 
For an exponential profile 
$h_\mathrm{eff}=z_\mathrm{s}$ and for an isolated isothermal sheet $\propto
\mathrm{sech}^2(z/2z_\mathrm{s})$ we find $h_\mathrm{eff}=2z_\mathrm{s}$. Most
realistic profiles are somewhere in-between.
From the normalized version of equation \ref{eqSigmas} we calculate 
$h_\mathrm{eff}/z_\mathrm{e}$.
The effective exponential scale height $z_\mathrm{s}/z_\mathrm{e}$ of the stellar disc  
is determined numerically by the mean exponential scale height in the range
 $z/z_\mathrm{e}=2\dots 5$.

The metallicity $\mathrm{[Fe/H]}$ affects the lifetimes, luminosities and colours 
of the stars and as a consequence also the mass loss of the
sub-populations. 
In order to account for the influence of the metal enrichment
we include a metal enrichment law AMR
that leads to
 a local metallicity distribution of late G dwarfs consistent with the
 observations. The properties of the stars and the
 stellar sub-populations are determined by population synthesis calculations (see
 Sect.\ \ref{pop}). 

The properties of MS stars with lifetime $\tau$ are determined by an
appropriate weighted average over the age range. We use
\bqn
S_\tau&=&\int_0^{\tau}\mathrm{SFR}(t')\dd \tau'\\
\rho_\tau(z)&=&\int_0^{\tau}\frac{\mathrm{SFR}(t')}{2h_\mathrm{age}(\tau')}
	\exp\left( \frac{-\Phi(z)}{\sigma^2_\mathrm{W}(\tau')}\right)\dd \tau' 
\eqn
where $S_\tau$ are all stars born up to age $\tau$ and $\rho_\tau(z)$ 
would be the density profile
of these stars, if mass loss by stellar evolution is ignored.
The thickness $h_\mathrm{ms}$ and
the normalised density profile $\rho_\mathrm{ms}(z)/\rho_\mathrm{ms}(0)$ 
of MS stars are given by
\bqn
h_\mathrm{ms}=\frac{S_\tau}{2\rho_\tau(0)}
&\mathrm{and}&
\frac{\rho_\mathrm{ms}(z)}{\rho_\mathrm{ms}(0)}=
\frac{\rho_\tau(z)}{\rho_\tau(0)}
\eqn
The normalised density profile is independent of the IMF. For the determination
of the number density profile of MS stars with lifetime $\tau$ the fraction of
stars $\Delta N$ in the corresponding mass interval of the IMF must be taken 
into account (similar for the mass density profile).

In a similar way we calculate the velocity dispersion profile
 $\sigma_\mathrm{ms}(z)$ and the velocity
distribution functions $f_\mathrm{ms}(W,z)$ for the MS stars using
\bqn
\sigma^2_\mathrm{ms}(z)&=&\frac{1}{\rho_\tau(z)}
\int_0^{\tau}\sigma^2_\mathrm{W}(\tau')\rho_{\tau'}(z)\dd \tau' \\
f_\mathrm{ms}(W,z)&=&\frac{1}{\rho_\tau(z)}\int_0^{\tau}
\frac{\rho_{\tau'}(z)}{\sigma_\mathrm{W}(\tau')\sqrt{2\pi}}
\exp\left( \frac{-W^2}{2\sigma^2_\mathrm{W}(\tau')}\right)\dd \tau'\qquad
\eqn

For the thick disc component we use a simple isothermal component adopting an age
larger than $t_\mathrm{p}$. The thick disc is parametrised by the surface
density $\Sigma_\mathrm{t}$ and the velocity dispersion
$\sigma_\mathrm{t}$. The density profile is calculated
self-consistently as for the thin disc sub-populations. The thick disc contributes
only to the velocity distribution functions of the lower MS with lifetimes larger
than $t_\mathrm{p}$. For the relative contribution to the $f_\mathrm{i}(W)$ we add
a correction factor ($\sim 1.8$) to the local mass ratio of thick and thin disc
to account for the different age distribution of thin and thick disc. The larger
mean age of the thick disc leads to
an enhanced number density of low mass stars due 
to the larger mass loss by stellar evolution.

\subsection{Gas component \label{gas}}

For the gas component we use a simple model to account for the
gravitational potential of the gas.
The vertical profile of the gas component that is used for the gravitational 
force of the gas, is constructed dynamically 
like the stellar component. The gas distribution is modelled by
distributing the gas with a constant rate up to a maximum age
over the  velocity
dispersion range $s_\mathrm{g}\sigma_\mathrm{W}(\tau)$ of the young stars 
reduced by a factor $s_\mathrm{g}$ to account for a smaller minimum value. 
By varying these parameters the peakyness and the width of the 
gas density profile can be
changed. We adjust the (half-)thickness and the exponential scale-height of the 
gas component to the observed values of $h_\mathrm{eff,g} \approx 150$\,pc 
and $z_\mathrm{g}\approx100$\,pc \citep{dic90}. 
We determine $z_\mathrm{g}$ numerically in a similar way as $z_\mathrm{s}$ but
closer to the mid-plane in the range of $z/(0.3\,z_\mathrm{e})=2\dots 5$.
The surface density $\Sigma_\mathrm{g}$ of the gas is iterated to reproduce the
observed value.

\subsection{Dark matter halo \label{halo}}

The halo does not fulfil the thin disc approximation. For a spherical halo we
get the vertical component of the force to lowest order from
\bq
\frac{\dd\Phi_\mathrm{h}}{\dd z}= \frac{v_\mathrm{c,h}^2}{R^2}z\quad, 
\eq
with $r^2=R^2+z^2$ and circular velocity $v_\mathrm{c,h}$ of the DM halo
at radius $R$. 
Comparing this with the one-dimensional Poisson equation from the thin disc
approximation (integrated over $z$ 
near the mid-plane to lowest order for small $z$)
\bq
\frac{\dd\Phi}{\dd z}\approx 4\pi G \rho_0 z
\eq
we can use for the local halo density 
\bq
\rho_\mathrm{h0}=\frac{v_\mathrm{c,h}^2}{4\pi G R^2}
\eq
to be consistent with the spherical distribution. 
Only for a singular isothermal sphere this virtual halo density corresponds 
exactly to the real local DM density. 
The second parameter is the halo velocity dispersion 
$\sigma_\mathrm{h}$. Comparing the spherical distribution of the halo density
and the vertical profile of a self-gravitating isothermal sheet up to 
second order in $z$ we find the relation
\bq
\sigma_\mathrm{h}^2=\frac{v_\mathrm{c,h}^2}
{\dd \ln \rho_\mathrm{h}/\dd \ln R}
\eq
For an isothermal sphere the logarithmic derivative of the density equals two.
As a consequence we can apply the thin disc approximation also for the halo 
by using the virtual halo density $\rho_\mathrm{h0}$ 
and velocity dispersion $\sigma_\mathrm{h}$ estimated from the 
circular velocity $v_\mathrm{c,h}$.
For a singular isothermal sphere these quantities correspond to the physical 
values. The effect of different radial profiles,
anisotropy and flattening would lead to correction factors. 

The DM halo is parametrised as an isothermal component similar to the thick disc 
by the velocity dispersion $\sigma_\mathrm{h}$ and the surface
density $\Sigma_\mathrm{h}$, which is implicitly determined by the total 
surface density.


\subsection{Stellar population synthesis \label{pop}}

For the determination of luminosities, main sequence lifetimes and mass loss we
are not interpolating directly evolutionary tracks of a set of stellar masses
and metallicities. In order to get a complete coverage of stellar masses we use
instead the stellar population synthesis code PEGASE 
\citep{pegase} to calculate mass loss and luminosities
of ``pseudo'' simple stellar 
populations (SSP). This means that the PEGASE code 
is used to calculate the
integrated luminosities and colour indices for a stellar population 
created in a single star-burst as function of age. 
These SSPs are then used to assemble
a stellar population with a given star formation history, in the sense that
the star formation history is assembled by a series of star-bursts. 
In this way we can assemble stellar populations for varying SFR and metal
enrichment $\mathrm{[Fe/H]}$. 
Our SSPs are modelled by a constant star formation rate with a
duration of 25\,Myr, the time resolution of the disc model. 
This is done
for a set of different metallicities and intermediate values from the chemical 
enrichment are modelled by linear interpolation.

\begin{figure}[t]
\centerline{\resizebox{0.98\hsize}{!}{\includegraphics[angle=270]{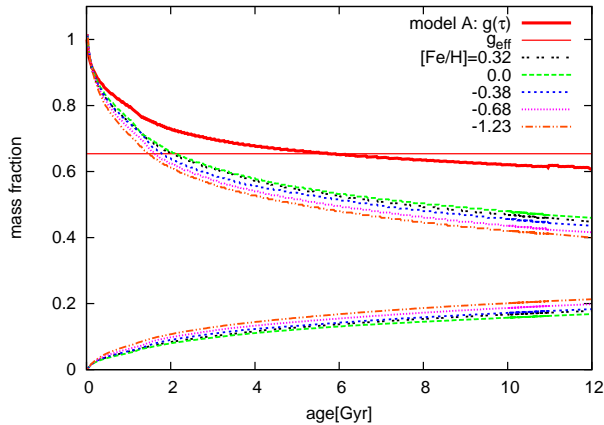}}}
\caption[]{
Mass loss due to stellar evolution. The two sets of thin lines show the fractions 
of luminous mass and of remnants (upper and lower curves, respectively)
 for the set of metallicities used in the PEGASE code. The
full (red) thick line shows the total stellar mass fraction (luminous matter plus remnants)
 of the fiducial model A taking into account the age-metallicity relation.
The overall present day mass fraction of the stars is
 given by $g_\mathrm{eff}$.
  }
\label{figmassloss}
\end{figure}

The application of the PEGASE code is twofold. On the one hand mass
loss due to stellar winds, planetary nebulae and supernovae determines 
$g(\tau)$, the mass fraction remaining in the stellar component as a
function of age $\tau$. This depends on the IMF and the metallicity.
We adopt a Scalo-like IMF \citep{sca86} given by
\bqn
\dd N &\propto& M^{\alpha}\dd M \label{eqscalo}\\
&& \alpha=\left\{\begin{array}{lcc}
        -1.25&&0.08\le M/\msun < 1\\
        -2.35&for&1\le M/\msun < 2\\
        -3.0&&2\le M/\msun < 100
        \end{array}\right.\quad .
\eqn
and five different metallicities 
([Fe/H]= -1.23, -0.68, -0.38, 0.0, 0.32), 
which are
the input parameters of the code. Fig.\ \ref{figmassloss} shows the
mass loss for the different metallicities. The fractions of luminous matter and 
of stellar remnants for the set of input 
metallicities are shown separately. The sum of both contributions for each metallicity vary by a few
percent only and are therefore not shown. The thick full (red) line
shows for the fiducial model A the fraction of stellar 
mass $g(\tau)$ in the present day stellar
disc as a function of age including the chemical evolution. 
For the oldest age-bins about 40\% of
the stellar mass is lost by stellar evolution. The total
fraction of stellar mass $g_\mathrm{eff}=0.654$ is indicated by the horizontal
line.

\begin{figure}[t]
\centerline{\resizebox{0.98\hsize}{!}{\includegraphics[angle=0]{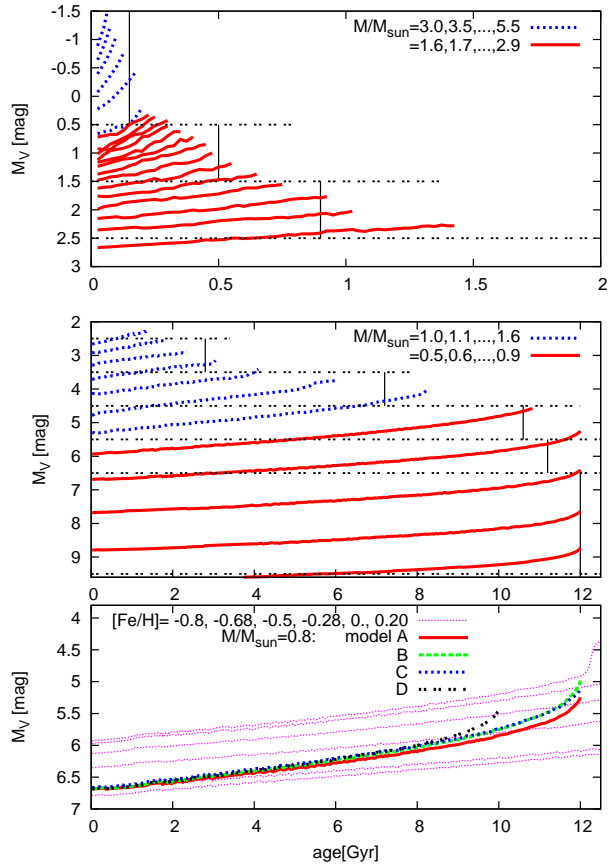}}}
\caption[]{
The lower panel shows the age dependence of the absolute luminosity 
$M_\mathrm{V}$ for stars with $M=0.8\,\msun$ (averaged over $0.1\,\msun$). Thin
lines are the evolution for different metallicities 
$\mathrm{[Fe/H]}=-0.8; -0.68; -0.5; -0.28; 0.0; 0.20$ and the thick line 
shows the
present day luminosity of the stars as a function of age taking into account
the age-metallicity relation.
The upper panels show the luminosity of mass bins along the main sequence as a
function of age taking the chemical enrichment into account. Vertical lines give
the adopted lifetimes for each luminosity interval.
}
\label{figvage}
\end{figure}

In the second application we determine the MS lifetimes and
luminosities for stellar mass bins.
Here we use the PEGASE code to compute V-band luminosities of isochrones in
small mass bins.  These are needed to estimate the mean
MS lifetime as a 
function of the absolute V-band luminosity $M_\mathrm{V}$ for the calculation of the
velocity distribution functions $f(|W|)$ of these stars.
We apply the PEGASE code to piecewise constant IMFs for mass-bins with
$0.1\msun$ and for a finer grid of metallicities 
([Fe/H]= -0.8, -0.68, -0.5, -0.28, 0.0, 0.20, 0.32). 
In Fig.\ \ref{figvage} we show $M_\mathrm{V}$ for the different mass bins as a
function of age. The lower panel gives the luminosity evolution 
of a $M=0.8\,\msun$ star
for the different metallicities (thin lines) and the age dependence of
the luminosity for different disc models for $M=0.8\,\msun$ demonstrating the significant
variation over the whole age range. The upper panels show the age dependent
luminosities for all mass bins. These are again not the luminosity
evolutions of the stars but the present day properties of the stellar population
taking into account the age-metallicity relation. The vertical lines indicate 
the estimated mean
MS lifetimes in the corresponding luminosity bins. 

For a consistent model the locations of
the mass tracks $(B-V,M_\mathrm{V})(\tau)$ in the Hertzsprung-Russell diagram
 (HR-diagram) should cover the same area used for the selection of main sequence stars
(see Fig.\ \ref{fighr}). Further information on the selection criteria are
given in Sect.\ \ref{observ}. The effect of choosing
these lifetimes on the determination of 
the velocity distribution functions $f(|W|)$ in the magnitude bins are discussed
in Sect.\ \ref{kinematics}.

\section{Observational Data} \label{observ}

In the model we assume that each stellar sub-population is in dynamical
equilibrium in the vertical direction. For the high luminosity end of the main
sequence with average age
 below 1\,Gyr this is not guaranteed. On the other hand the observed velocity
 distribution functions and also the vertical density profile of A stars
 measured by \citet{hol00} are hints for equilibrium
 distributions. Additionally we use large radii of 200\,pc for these sub-samples
 and determine a spatially averaged distribution. All sub-samples
 can be used as independent samples to determine the total vertical 
gravitational potential $\phi(z)$. 
The velocity distribution functions $f(|W|)$ of the samples depend on the
age distribution of the stars. Therefore we use main sequence stars, where
$f(|W|)$ is a function of the lifetime of the stars. In order to determine the
vertical velocity distribution function $f(|W|)$ in the solar neighbourhood we
need kinematically unbiased samples with space velocities. 

For the determination of the AVR we use only the measured velocity dispersions
in the magnitude bins along the main sequence using stellar lifetimes.
An independent stellar sub-sample are the
McCormick K and M dwarfs \citep{vys63}. Detected by a spectroscopic survey  
they are free from kinematic bias. Altogether 516 stars show reliable distances - almost
all from the Hipparcos Catalogue - and space velocity
components. All stars within the 25\,pc sphere are also used to determine 
independently the velocity distribution
function $f(|W|)$ of stars with lifetime larger than 12\,Gyr 
(see  Fig.\ \ref{figfw}). 
For a sub-sample of about 300 stars \citet{wil70} estimated the 
CaII emission intensity at the H
and K lines in a relative scale allowing to construct six different age bins 
under the assumption of
a constant SFR (see Jahrei{\ss} and Wielen, 1997).    
For each bin the vertical velocity dispersion is determined. We do not use these
sub-samples in our model for two reasons. The determination of the mean ages 
of the bins are inconsistent with our disc model and for the velocity
distribution functions in the age bins the number of stars
is too low to obtain reliable $f(|W|)(t)$. The AVR of this sample
 is shown in the lower panel of Fig.\ \ref{figsig} only for
comparison.
 
\begin{figure}
\centerline{\resizebox{0.98\hsize}{!}{\includegraphics[angle=0]{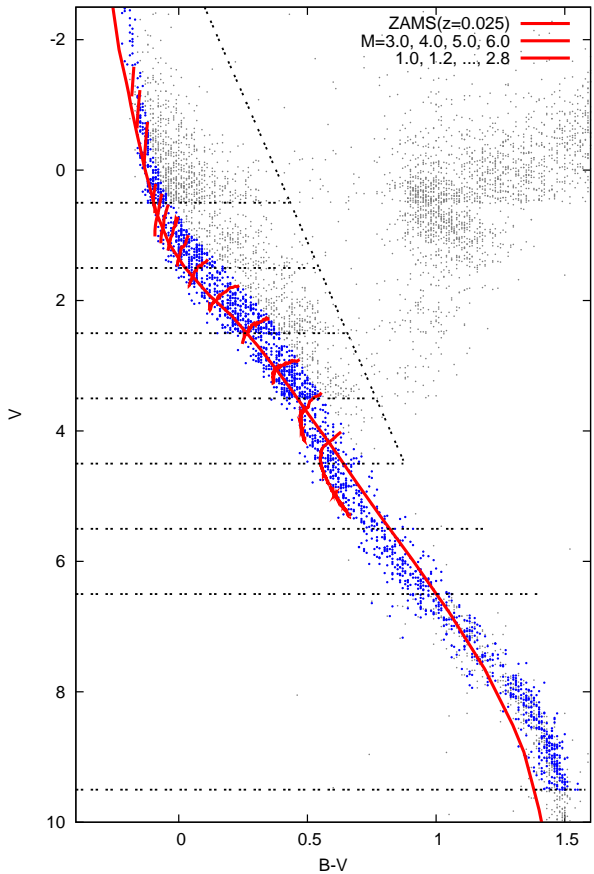}}}
\caption[]{The HR-diagram shows all Hipparcos stars (grey dots) 
with $\sigma_{\pi}/\pi \le 0.15 $
within the distance limits chosen (see column 3 in Table 1)
The selected main sequence 
stars are over-plotted by larger dark dots. The eight
magnitude bins are indicated by the horizontal dotted lines and the inclined
line marks the selection including the turnoff stars.
The full lines show the ZAMS with present day metallicity and the 'tracks'
from Fig. \ref{figvage} used for the selection process.
}
\label{fighr}
\end{figure}
For the determination of $f(|W|)$ along the main sequence we use the Hipparcos
stars with good space velocities supplemented at the faint end down to 
$M_\mathrm{V}=9.5$\,mag by stars from an updated version of the 
Catalogue of Nearby stars (CNS4) \citep{jah97}, i.e. also most of 
the CNS4 data rely on Hipparcos results. 
 For the determination of the space velocities good distances, proper
motions and radial velocities are required. This was achieved in combining the Hipparcos
data with radial velocities originating from an unpublished compilation of the "best" radial 
velocities for the nearby stars that was then extended to all Hipparcos stars.

We apply a 2-step selection process. In the first step the absolute visual
 magnitude $ M_V $ of the Hipparcos stars was
determined from the visual magnitude and the trigonometric parallax given in the Hipparcos
catalogue. Only stars with $\sigma_{\pi}/\pi$ smaller than 15\% were taken into
account. In the case of binaries resolved by Hipparcos the $ M_V $ of the brighter component
 was used. It was
calculated from the combined magnitude and the magnitude difference measured by Hipparcos.
The stellar sample is divided into magnitude bins 
$M_\mathrm{V} = -1\pm1.5, 1\pm0.5, ..., 6\pm0.5, 8\pm1.5$\,mag.
In order to avoid a kinematic bias we restrict the distances of the stars in
each magnitude bin to be well within the completeness limit of the Hipparcos 
catalogue determined by the magnitude limit $ V \sim 7.3$\,mag  
of the Hipparcos Survey. The faintest magnitude bin is restricted
to a distance of 25\,pc relying on the completeness of the CNS4 
(grey dots in Fig.\ \ref{fighr}).

In the second step we select in the HR-diagram 
(Fig.\ \ref{fighr}) a regime along the main sequence in order to exclude
most of the turnoff stars as well as all giants and white dwarfs.
As zero age main sequence we used the $(B-V, M_V)$-relation for the present-day
metallicity z=0.025
originating from the Padua-isochrones \citep{ber94,gir02}
(URL: \mbox{http://stev.oapd.inaf.it/$\sim$lgirardi/cgi-bin/cmd}). Then 
all stars  in the
magnitude range $M_\mathrm{V,ms}\pm 0.8$\,mag were selected. The reason for
this additional selection is to match the mass tracks from stellar population
synthesis up to the turnoff point described in Sect.\ \ref{pop}. A selection of
these tracks are over-plotted in the HR-diagram.
 For the brightest magnitude bin 
the ZAMS is almost vertical and the stars  
redder than $(B-V)_{ZAMS} + 0.10$ were excluded.

This selection is chosen to 
include the metal-poor MS 
stars below the ZAMS and to allow for a metallicity scatter of about 0.2dex
as well as to include unresolved MS binaries above the
single star MS. On the other hand most turnoff stars are excluded,
because they would lead to an additional spread of stellar masses and lifetimes 
in the magnitude bins. For a consistency
check we analysed a second set of stellar samples by including the turnoff stars
bluer than the inclined line in (Fig.\ \ref{fighr}) and using the corresponding
larger lifetimes derived from stellar evolution. 
 The properties of the resulting samples of main sequence stars are
collected in Table\ \ref{tabdata}.

\begin{table}
\caption {Complete samples of nearby main sequence stars:
 Column 1 lists the source catalogue. 
Column 2, 3 and 4 list  the range
 in absolute visual magnitude, the selected distance limit, and the 
 total number of stars available, respectively.
 Column 5 and 6 list the number of
stars removed due to poor parallaxes ($\sigma_{\pi}/\pi > .15 $) and 
unknown or poor radial 
velocities. In column 7 the
remaining number of main sequence stars with good space velocity components is 
given and the last column contains the corresponding number of stars including
the turnoff stars.}
\begin{tabular}{l@{\extracolsep\fill}rrrrrrr}
\hline 
source & $M_{V}$  & $d_{lim}$ &  N & n$^*$ & no RV & N$_{fin}$ & $N_{to}$\\
      & [mag]   & [pc]      &    &         &         &           \\
\hline 
Hip & -2, -1, 0 & 200 &    304 &   39 &  15   & 250 & 931\\
Hip &  1 & 100      &   242 &    0 &   1   & 241 & 497\\
Hip &  2 &  75      &   401 &    8 &   7   & 386 & 650\\
Hip &  3 &  50      &   352 &    0 &   4   & 348 & 491\\
Hip &  4 &  30      &   172 &    2 &   0   & 170 & 194\\
Hip &  5 &  25      &   172 &    0 &   0   & 172 \\
Hip &  6 &  25      &   170 &    0 &   2   & 168 \\
CNS & 7, 8, 9  &  25  &   525 &      &  84   & 441 \\
\hline
\end{tabular}
\label{tabdata}
\end{table}

For the determination of the velocity distribution functions $f(|W|)$ we correct
for the peculiar motion of the Sun using 
the local standard of rest  value $W_\odot=7$\,km/s given in \citet{del65} that is in close 
agreement to the modern  $W_\odot=7.17 \pm 0.38$\,km/s determined by 
\citet{deh98} from the Hipparcos data.
The resulting
normalised distribution functions are shown in Fig.\ \ref{figfw} 
with a binning of 5\,km/s in $|W|$.
For the determination of the velocity dispersions we exclude
stars with high velocities ($|W|>40\,km/s$ for $M_{\mathrm V}<3.5$ and
$|W|>60\,km/s$ for $M_{\mathrm V}>3.5$). 

\section{Properties of the disc} \label{disc}

We discuss first the algorithm to determine the parameters of best fitting
models. Then
we describe our fiducial model A in great detail and compare the properties with
a selection of other models of similar fitting quality.

\subsection{Best fit procedure \label{fit}}

The disc model depends on a set of input functions and parameters with strong
correlations and degeneracies. Thus a simultaneous best fit of all parameters and
an automatic selection of the astrophysical best model is impossible.
Therefore it is very useful to sort the input
quantities hierarchically by there impact on the velocity distribution
functions of MS stars and the structure of the disc.

\begin{figure}
\centerline{\resizebox{0.98\hsize}{!}{\includegraphics[angle=0]{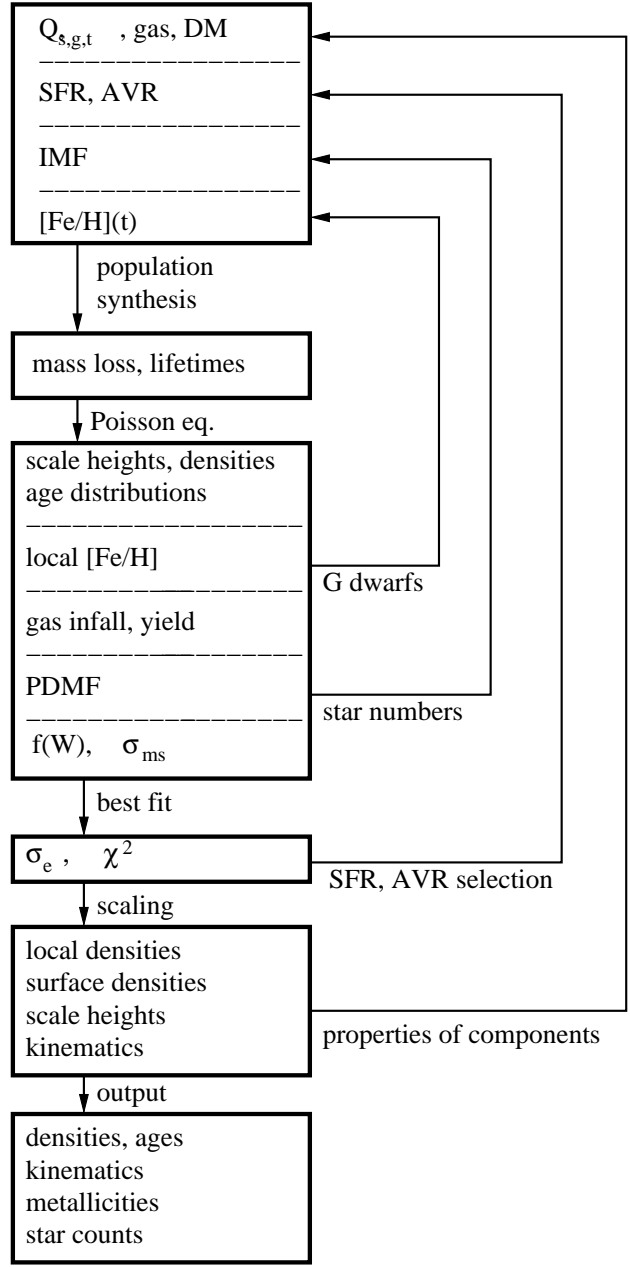}}}
\caption[]{
Flowchart of fitting procedure. For a description see text.
}
\label{figfit}
\end{figure}
The fitting procedure can be split into three parts: 1) pre-processing to calculate
a normalised model; 2) best fit of $f_\mathrm{i}(|W|)$ and scaling of the
model; 3) post-processing to calculate the output and predictions (see figure
\ref{figfit}). We start with a reasonable set of mass fractions $Q_\mathrm{s,g,t}$
of the components including the kinematics of the gas, thick disc and DM halo
and fix the IMF.
From large sets of template functions for the normalised SFR and AVR we
select up to ten of each.
We select for each SFR a metal enrichment law.
Then we construct models for all (SFR,AVR) combinations by calculating the mass
loss due to stellar evolution, solving the Poisson equation and determining
the local stellar age distribution. With pre-calculated MS lifetimes we
determine the normalised velocity distribution functions $f_\mathrm{i}(|W|)$ 
for all nine stellar sub-samples. 

In each model we determine 
the common scaling factor $\sigma_\mathrm{e}$ for all  
$f_\mathrm{i}(|W|)$ by a nonlinear Levenberg-Marquardt best fit algorithm
\citep{pre92}. It turned out that a standard
$\chi^2$ fit is very sensitive to the choice of the velocity range in $|W|$ due to
the sparse population in the higher velocity bins. Therefore we tested the
$Z^2$ statistics from \citet{luc00} which is a modified $\chi^2$
statistics leading to a standard
distribution also for sparse data sets. This algorithm is very stable and we
fix the velocity range to $|W|<40$\,km/s for the bright bins with $M_V<3.5$ and
60\,km/s for the faint bins. For 90 degrees of freedom a reduced $\chi^2=1.2$
corresponds to a 10\% level of significance that the data are consistent 
with the model. An inspection of the individual reduced $\chi^2$ for the
$f_\mathrm{i}(|W|)$ shows that the brightest bin with $\chi^2>3.0$ is 
inconsistent with the model on the 99.5\% level for all models. This result
confirms that these very  young stars are not in dynamical equilibrium.
Nevertheless we include this bin in our calculation in order not to ignore the
low velocity dispersion in that bin. Tests by excluding the first bin and  by
fitting also the normalisations $n_\mathrm{i}$ of the $f_\mathrm{i}(|W|)$ does 
not change the results significantly.

For further investigations we consider only models with $\chi^2<1.2$ which
already leads to strong constraints in the allowed heating laws AVR. All
further iterations are done for individual pairs (SFR,AVR). In the next step
the calculated G dwarf metallicity is compared to the observations and a new
chemical enrichment law is constructed to give a reasonable fit. Corrections
via stellar mass loss and MS lifetimes are small and the new model can be used
to derive the conversion factor from the IMF and the local present day mass 
function (PDMF). Now the IMF can be determined from the observed PDMF which is
done in the current phase only for comparison and not iterated.

For the next iteration step we need to scale the model to physical quantities.
The first of the two free scaling factors is $\sigma_\mathrm{e}$ which is 
already fixed by the best fit procedure.  
For the second parameter we decided to fix the local stellar density 
including the thick disc contribution to
$\rho_\mathrm{s0}+\rho_\mathrm{t0}=0.039\,\msun\,pc^{-3}$,
which is the best observed quantity for the local disc model 
\citep{jah97}. 
We can rewrite equation \ref{eqze} to derive $z_\mathrm{e}$ by
\bqn
z_\mathrm{e}^2=\frac{Q_\mathrm{s}}{4\pi G C_\mathrm{eff}}
	\frac{\sigma_\mathrm{e}^2}{\rho_\mathrm{s0}}&\mathrm{with}&
	C_\mathrm{eff}=h_\mathrm{eff}/z_\mathrm{e}
\label{eqzescale}
\eqn
Then $h_\mathrm{eff}$ can be scaled and the total surface density is
determined by
\bq
\Sigma_\mathrm{tot}=\frac{2h_\mathrm{eff}\rho_\mathrm{s0}}{Q_\mathrm{s}}
\eq
All other quantities can be scaled in a similar way. 
In this final iteration the input parameters for the surface densities 
and the kinematics are adapted in such a way as to reproduce the known constraints on
surface densities, local densities and scale heights.
The thick disc parameters can be freely chosen in a wide range with very little
influence on the other disc properties.
The kinematic parameters of gas and DM halo are 
forced to get the correct gas scale height $z_\mathrm{g}\sim 100$\,pc 
and DM velocity dispersion $\sigma_\mathrm{h}\sim 140$\,km/s. 
The value of $Q_\mathrm{h}$ and as a consequence the local halo density 
$\rho_\mathrm{h0}$ are implicitly determined by fixing the other $Q$ values. 
The value of $Q_\mathrm{g}$ is chosen in a simple step
 to be consistent with the observed surface density of 
$\Sigma_\mathrm{g}\approx 10-13\,\msun pc^{-2}$ \citep{dic90}.
Adapting the thin disc parameter $Q_\mathrm{s}$ is more sophisticated,
because it has a significant influence on the shape of the gravitational potential.
For the boundary conditions we have a relatively large freedom and compare models
which fulfil different constraints on the thin disc scale height $z_\mathrm{s}$ or on
surface densities.

We discuss more details of parameter choice in the next subsections and a more
detailed investigation of the best fit procedure and model selection will be given
in Paper II.

In the next subsections we discuss the selection of the main input 
functions SFR and AVR and
other properties of the disc models including
predictions for future observations. 

\subsection{Star formation history and dynamical heating \label{sfr}}

The main input functions that are to be determined, are the SFR and AVR. 
Since
the SFR is a priori not strongly constrained, we start with template sets of
smooth functions of different analytic type: pure exponentials with different
disc age and decay timescale; linear increase combined
with exponential decline; algebraic with linear increase and polynomial decline.
For the heating function AVR we use templates of power laws with different power
law indices and offsets in time to allow a nonzero present day value. A second
set has a constant velocity dispersion for ages larger than 6 or 9\,Gyr as
proposed by \citet{fre91}. A third set of polynomials was chosen to
produce an extra  flattening or steepening in the AVR around intermediate age of 
$\sim 6$\,Gyr.

In a preliminary study we select those parameter regimes which result in a
reasonable $\chi^2<1.2$ for at least one heating function AVR. 
We found that the polynomial functions do not improve
the fits for any SFR significantly compared to corresponding power law AVRs.
The range of consistent SFRs, based on the local kinematic data alone, is weakly
constraint:
We find for the present day star formation rate
SFR$_\mathrm{p} > 1.2\msun/pc^2/Gyr$ 
(smoothed over the last 500\,Myr) and
for stars older than 10\,Gyr a contribution less than 25\%.
This means that the overall decline of the SFR is limited to a
factor of 4--5.

In the main step we restrict the models further to 
(SFR,AVR) pairs with $\chi^2<1.1$. All models with $\chi^2$ below this limit are 
consistent with the $f_\mathrm{i}(|W|)$ of the local MS stars. For each selected SFR
the possible AVRs are restricted to at most a few. Since there  is still
quite a variety of very different models left over, we restricted the further
investigation to a
representative selection of (SFR,AVR) combinations. The fine-tuning of the model
parameters (metal enrichment, gas and DM properties, surface density constraints) is
done only for these models. It turned out that these parameter adjustments do not
change the best fit significantly (similar $\chi^2$ and $\sigma_\mathrm{e}$). 
In the models we fix the surface density of the gas to the observed value and the stellar
surface density is determined by the local stellar density and the best fit 
$\sigma_\mathrm{e}$. Therefore
the DM surface density, i.e. the mass fraction $Q_\mathrm{h}$, is mainly 
determined by the surface density of all components
$\Sigma(|z|<1.1\,kpc)\sim 75\msun/pc^2$ \citep{kui91}, \citep{hol04}. 
It must be determined by an iteration process, because changing the $Q$'s also
changes the scale heights of all components. 
The second boundary condition is the scale height $z_\mathrm{s}$  of the thin
disc. It is connected to $Q_\mathrm{s}$ via the scaling length $z_\mathrm{e}$ 
(equation \ref{eqzescale}). 
The standard value of 325$\pm$50\,pc determined by \citet{bah84c} from star
counts is still reliable and widely used. Since this scale height is calculated
from a one component exponential thin disc, it corresponds to the mean
exponential falloff in the range of $|z|\sim 100 - 1000$\,pc which is much closer to
the mid-plane than our value $z_\mathrm{s}$ measuring the exponential scale height
in the range $|z|\sim 600 - 1500$\,pc. An inspection of the density profile
shapes (figure \ref{figrho}) shows that closer to the mid-plane the local 
scale height is $\sim$20\% larger than $z_\mathrm{s}$. Therefore we try to reach
in the models $z_\mathrm{s}\sim 270-300$\,pc which is not possible in all cases.
Reducing the DM fraction increases
the thin disc fraction $Q_\mathrm{s}$ and as a consequence the scale height 
$z_\mathrm{s}$. As a bottom line there is a trade-off to reach both constraints
given by $\Sigma(|z|<1.1\,\mathrm{kpc})$ and $z_\mathrm{s}$.
A general overview over all models and a detailed discussion of the parameter
dependence will be given in Paper II \citep{jus09}.

\begin{figure}
\centerline{\resizebox{0.98\hsize}{!}{\includegraphics[angle=0]{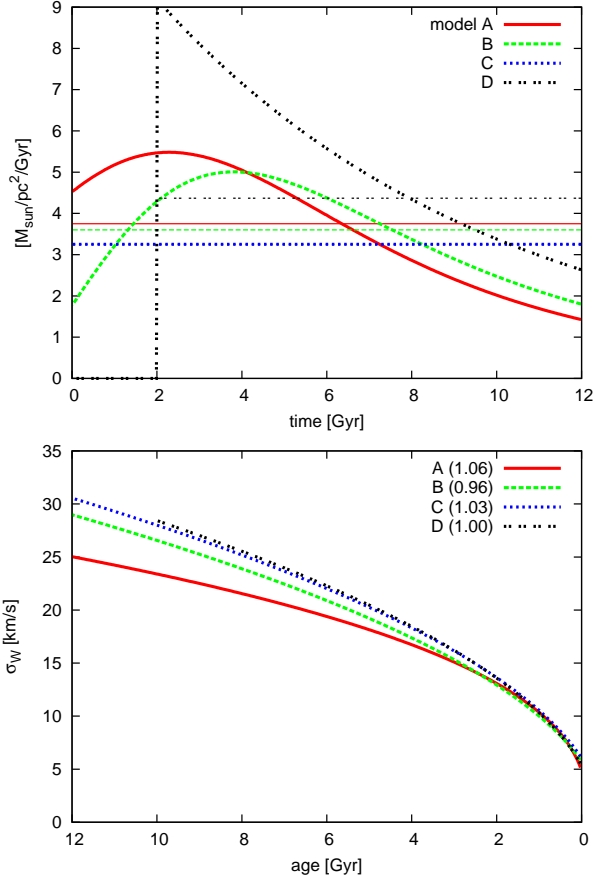}}}
\caption[]{
The upper panel shows the SFRs of models A--D. The horizontal lines are the
corresponding mean values $\langle \mathrm{SFR} \rangle\,$.
The lower panel gives the AVRs of models A--D($\chi^2$ values in
parentheses). 
Age is running backwards.
}
\label{figsfr}
\end{figure}

We present four models A--D with very different SFR. For clarity we use model A as
the fiducial model that will be described in great detail. The  SFR of this model
is very close to that investigated also by \citet{aum09}. We discuss the
similarities and differences of models B--D with respect to model A. 
The data of models E--G are added to show the effect of some parameter variations.
The basic properties of the models are
\begin{description}
\item[Model A:] Fiducial model with maximum SFR 10\,Gyr ago and mass fractions chosen to
get simultaneously a small thin disc scale height of $\sim 270$\,pc and a surface
density $\Sigma(|z|<1.1\,kpc)\sim 75\msun/pc^2$.
\item[Model B:] Model with smallest reduced $\chi^2=0.96$ of all models; optimised as
model A
\item[Model C:] Model with constant SFR; DM fraction increased to the upper limit in
order to get a reasonable thin disc scale height of $z_\mathrm{s} \sim 300$\,pc
\item[Model D:] Model with reduced disc age of $t_\mathrm{p}=10$\,Gyr and exponential
SFR with shortest consistent decline timescale.
\item[Model E:] Model with same SFR, AVR as model A but neglecting the thick
disc component.
\item[Model F:] Like model A but with larger local stellar density.
\item[Model G:] Like model F but with larger gas fraction in order to increase
the local density further.
\end{description}
The SFRs and AVRs of models A--D are shown in figure \ref{figsfr}
together with the mean SFR determined by $\langle SFR \rangle\, =S_0/t_\mathrm{p}$.
A list of model parameters and derived physical
quantities discussed in this paper is given in Table \ref{tabmod} 
together with the corresponding data of other authors.

\begin{table*}
\caption {Disc properties for models A--G and some models from the literature.
Global parameters:
$\chi^2$ value of the fit; maximum height $z_\mathrm{max}$, 
local density$\rho_\mathrm{0}$, surface densities below 0.35 and 1.1\,kpc,
total surface density of the disc.
Thin disc properties:
maximum velocity dispersion $\sigma_\mathrm{e}$,
mean and present day SFR, 
local density $\rho_\mathrm{s0}$;
surface density $\Sigma_\mathrm{s}$ up to $z_\mathrm{max}$,
(half-)thickness $h_\mathrm{eff}$,
exponential scale height $z_\mathrm{s}$;
Thick disc properties:
velocity dispersion $\sigma_\mathrm{t}$,
local density $\rho_\mathrm{t0}$;
surface density $\Sigma_\mathrm{t}$ up to $z_\mathrm{max}$,
exponential scale height $z_\mathrm{t}$;
power law index $\alpha_\mathrm{t}$ of sech$^{\alpha_\mathrm{t}}$ fit.
Gas component:
local density $\rho_\mathrm{g0}$;
surface density $\Sigma_\mathrm{g}$ up to $z_\mathrm{max}$,
(half-)thickness $h_\mathrm{eff,g}$;
exponential scale heights $z_\mathrm{g}$;
DM halo:
velocity dispersion $\sigma_\mathrm{h}$,
local density $\rho_\mathrm{h0}$;
surface density $\Sigma_\mathrm{h}$ up to $z_\mathrm{max}$.
}
\begin{center}
\begin{tabular}{ll|ccccccc|l}
\hline 
quantity &	unit			& A 	& B	& C	& D	& E	& F	& G	&other  sources\\
	&			& fiducial &min.$\chi^2$& const.SFR&lt.=10Gyr&no.thick d.&larger $\rho_{s0}$&more gas&\\
\hline 
red. $\chi^2$		& -		&1.057	&0.960  &1.025  &0.994  &1.046  &1.056	&1.054	&   \\
$z_\mathrm{max}$	& kpc		&2.3	&2.7	&2.7	&3.1	&2.3	&2.3	&2.3	&   \\
$\rho_\mathrm{0}$ 	& $\msun/pc^3$ 	&0.088	&0.091  &0.092  &0.092  &0.088  &0.094	&0.101	&0.076$^1$, 0.098$^2$   \\
$\Sigma(<0.35)$  	& $\msun/pc^2$	&41.1	&41.3   &42.3   &40.7   &41.0   &43.5	&44.5	&41$^3$   \\
$\Sigma(<1.1)$  	& "	 	&73.5	&73.2   &79.3   &71.8   &72.5   &74.9	&74.5	&74$\pm 6^3$, 71$\pm 6^4$  \\
$\Sigma_\mathrm{disc}$  & "  		&45.2	&45.7   &42.1   &46.7   &41.8   &49.2	&50.4	& 56$\pm 6^3$, 48$\pm 9^4$  \\
\hline 
$\langle SFR\rangle$ & $\msun/pc^2/Gyr$	&3.75	&3.60   &3.25   &4.38	&3.99   &4.09	&4.02	&   \\
SFR$_\mathrm{p}$ 	& " 		&1.43	&1.80	&3.33	&2.20	&1.55	&1.58	&1.42	&    \\
$\sigma_\mathrm{e}$  	& km/s	 	&25.0	&28.4   &29.9   &28.0   &25.1   &25.0	&25.1	&17.5$^1$	 \\
$\rho_\mathrm{s0}$ 	& $\msun/pc^3$	&0.037	&0.037  &0.037  &0.037  &0.039  &0.041	&0.041	&0.045$^1$, 0.044$^2$   \\
$\Sigma_\mathrm{s}$ 	& $\msun/pc^2$ 	&29.4 	&28.6   &26.3   &29.2   &31.4	&32.1	&31.6	& 34.4$^2$    \\
$h_\mathrm{eff}$ 	& pc  		&400	&389	&357	&398	&402	&386	&379	&	    \\
$z_\mathrm{s}$  	& " 	 	&274	&295    &303    &309    &277    &266	&265	&	    \\
\hline 
$\sigma_\mathrm{t}$ 	& km/s		&45.1	&45.4   &44.9   &44.9   &-	&45.1	&45.1	& 37$^3$	 \\
$\rho_\mathrm{t0}$ 	& $\msun/pc^3$	&0.0022	&0.0023 &0.0022 &0.0023 &-      &0.0025	&0.0024	& 0.007$^3$  \\
$\Sigma_\mathrm{t}$ 	& $\msun/pc^2$ 	& 5.3	& 5.7   & 5.4   & 5.8   &-      & 5.7	& 5.6	&   \\
$z_\mathrm{t}$ 		& pc		& 793	& 807	& 680	& 815	&-	& 808	& 827	&	\\
$\alpha_\mathrm{t}$	&  -	 	&-1.19	&-1.17	&-1.60	&-1.13	&-	&-1.07	&-1.00	&	\\
\hline 
$\rho_\mathrm{g0}$ 	& $\msun/pc^3$	&0.035	&0.038  &0.035  &0.040  &0.034  &0.037	&0.042	&0.021$^1$, 0.050$^2$    \\
$\Sigma_\mathrm{g}$ 	& $\msun/pc^2$ 	&10.5	&11.4   &10.5   &11.7   &10.5   &11.4	&13.2	&6$^1$, 13$^2$   \\
$h_\mathrm{eff,g}$  	& pc	 	&151	&149    &151    &145    &151	&152	&148	&140$^1$	\\
$z_\mathrm{g}$  	& "	 	&94	&98	&98	&97	&94	&96	&94	&140$^1$	\\
\hline 
$\sigma_\mathrm{h}$  	& km/s	 	&140	&141    &141    &141    &141    &140	&140	&85$^1$   \\
$\rho_\mathrm{h0}$ 	& $\msun/pc^3$	&0.014	&0.014  &0.018  &0.013  &0.015  &0.013	&0.012	&0.01$^1$   \\
$\Sigma_\mathrm{h}$ 	& $\msun/pc^2$ 	&59.9	&68.6   &89.4   &70.1   &62.7   &54.4	&51.4	&   \\
\hline
\end{tabular}
\end{center}
$^1$ \citet{rob03}; 
$^2$ \citet{hol00}; 
$^3$ \citet{hol04}; 
$^4$ \citet{kui91}
\label{tabmod}
\end{table*}

For models A--D we give explicitly the input functions to allow the reader to
make comparison calculations.
We find for models A and B
\bqn
\mathrm{SFR}(t)&=&\langle \mathrm{SFR} \rangle\,\frac{(t+t_0)t_n^3}{(t^2+t_1^2)^2}\quad\mathrm{with}\\
A&:&\,t_0=5.6\,\mathrm{Gyr}\,;\, t_1=8.2\,\mathrm{Gyr}\,;\, t_n=9.9\,\mathrm{Gyr} \nonumber\\ 
B&:&\,t_0=1.13\,\mathrm{Gyr}\,;\, t_1=7.8\,\mathrm{Gyr}\,;\, t_n=11.7\,\mathrm{Gyr}\, , \nonumber
\eqn
for model C
\bqn
 \mathrm{SFR}(t)&=&\langle \mathrm{SFR} \rangle
\eqn
and for model D
\bqn
\mathrm{SFR}(t)&=&\langle \mathrm{SFR} \rangle\,\frac{t_\mathrm{p}}{t_\mathrm{e}}
	\frac{\exp(-t/t_\mathrm{e})}{1-\exp(-t_\mathrm{p}/t_\mathrm{e})}\\
D&:&\,t_\mathrm{e}=8\,\mathrm{Gyr}\,;\, t_\mathrm{p}=10\,\mathrm{Gyr}\, .\nonumber
\eqn
For the dynamical heating function AVR we use a power law
\bqn
\sigma_\mathrm{W}(\tau)&=&\sigma_\mathrm{e}
 \left(\frac{\tau+\tau_0}{t_\mathrm{p}+\tau_0}\right)^\alpha\quad\mathrm{with}\\
A&:&\,\alpha=0.375\,;\, \tau_0=0.17\,\mathrm{Gyr}\,;\,t_\mathrm{p}=12\,\mathrm{Gyr}\,  \nonumber\\
B&:&\,\alpha=0.5\,;\, \tau_0=0.5\,\mathrm{Gyr}\,;\,t_\mathrm{p}=12\,\mathrm{Gyr}\,  \nonumber\\
C&:&\,\alpha=0.5\,;\, \tau_0=0.5\,\mathrm{Gyr}\,;\,t_\mathrm{p}=12\,\mathrm{Gyr}\,  \nonumber\\
D&:&\,\alpha=0.5\,;\, \tau_0=0.32\,\mathrm{Gyr}\,;\,t_\mathrm{p}=10\,\mathrm{Gyr}\, . \nonumber
\eqn

Model C with constant SFR is of special interest, because it is still used in
many applications directly or implicitly. Model C fits well the local
kinematic data. Due to the large fraction of young and dynamically cool stars a
larger initial velocity dispersion and a higher fraction of DM matter is
needed. The relatively large values for $\Sigma(|z|<1.1\,\mathrm{kpc})$ and 
$z_\mathrm{s}$ are still within the limits.

\subsection{Kinematics of the disc \label{kinematics}}

\begin{figure*}
\centerline{\resizebox{0.7\hsize}{!}{\includegraphics[angle=0]{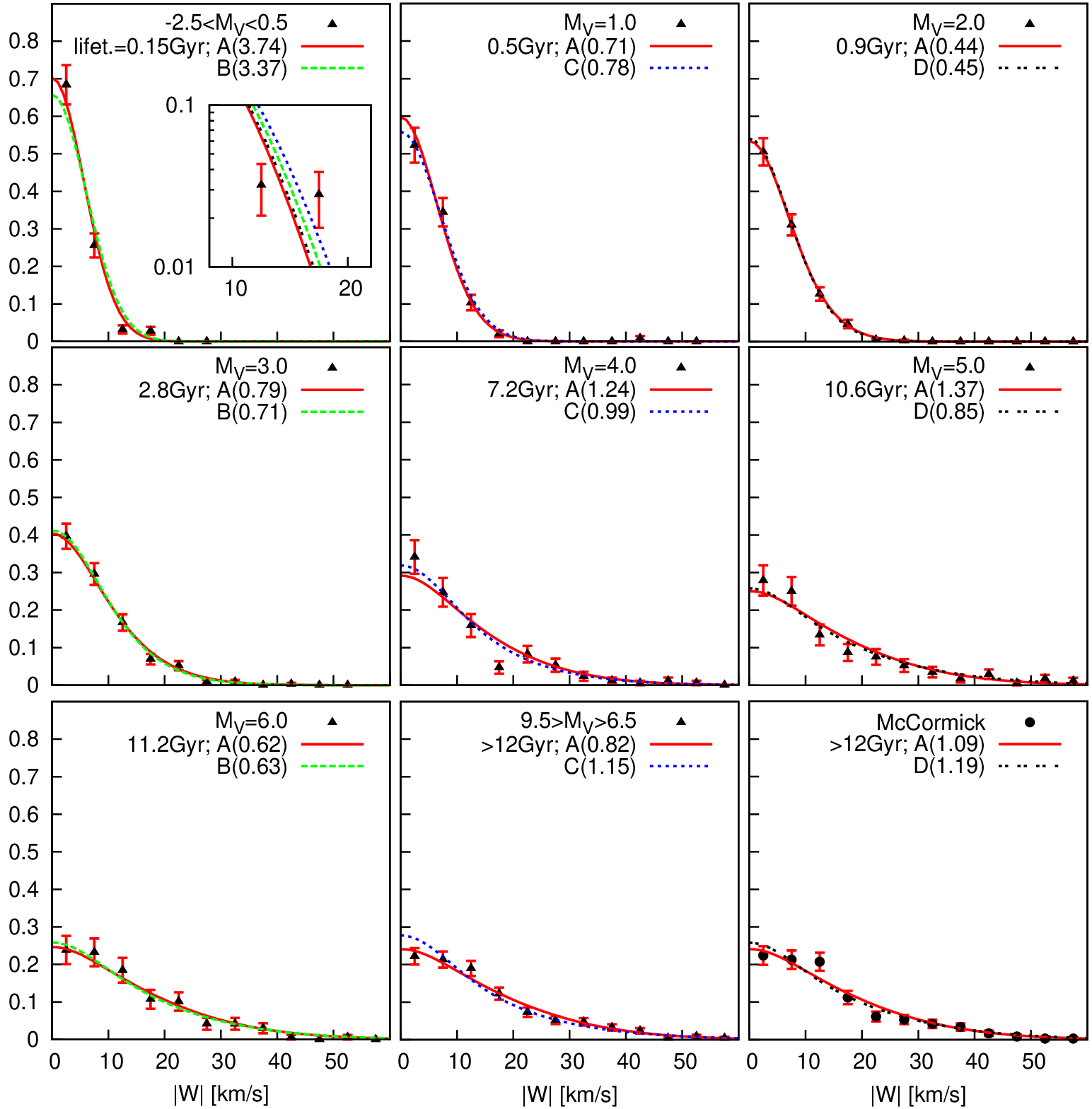}}}
\caption[]{The normalised velocity distribution functions $f_\mathrm{i}(|W|)$ for the 
different sub-samples given in Table \ref{tabdata} and for the McCormick stars 
(lower right). Symbols are the data and full lines represent the 
fiducial model A.
In each column the distribution functions of one of the comparison models are
added: B: best $\chi^2$; C: const. SFR; D: exp.SFR and lifetime=10\,Gyr.
The inset in the upper left panel shows the part of $f_\mathrm{i}(|W|)$ (in
log-scale), where the significant deviations occur.
}
\label{figfw}
\end{figure*}

\begin{figure}
\centerline{\resizebox{0.98\hsize}{!}{\includegraphics[angle=270]{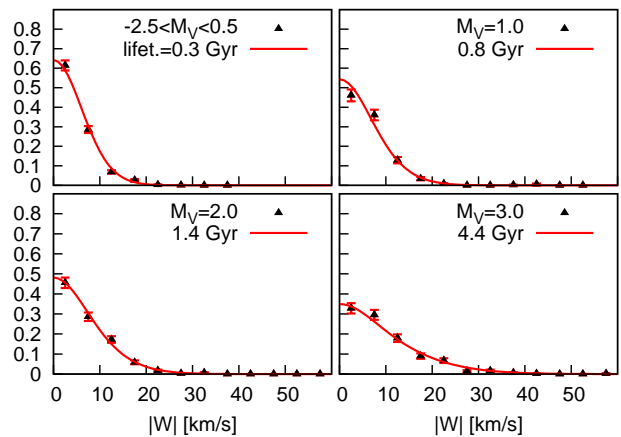}}}
\caption[]{The normalised velocity distribution functions $f(|W|)$ of the four
brightest bins as in Fig.
\ref{figfw} but with turnoff stars included.
}
\label{figfw2}
\end{figure}

All presented models show reasonable fits to the set of normalised 
velocity distribution functions $f_\mathrm{i}(|W|)$ for the nine samples of MS
stars described in section \ref{observ}, because the total reduced $\chi^2<1.1$. 
For models A--D the total $\chi^2$ values are given in the lower panel of
figure \ref{figsfr}.
In figure \ref{figfw} the data are compared to model A in all samples. For
comparison in each column one of the other models B--D are shown. 
 We also computed the $\chi^2_\mathrm{i}$ of the 
individual distribution functions $f_\mathrm{i}(|W|)$ to check the individual
contributions. The values are given in parentheses in figure \ref{figfw}.
The distribution of the $\chi^2_\mathrm{i}$ are consistent with a random
distribution with one exception.
In all models the
brightest bin has an individual reduced $\chi_\mathrm{1}^2>3$
for eight degrees of freedom (= number of bins). The inset in the upper left
panel of  figure \ref{figfw} shows the the distribution function of models A--D
in log-scale for the bins between 10 and 20\,km/s, where the significant
deviations occur. The $1-\sigma$ standard Poisson noise of the data (shown by
the errorbars) are not a precise measure of the contribution to 
$\chi^2$, since we use the $Z^2$ statistics (for details see Paper II).
This is a hint to a
significant departure from the assumed equilibrium distribution. 
This is not surprising, since the lifetime of
these stars is of the order of the vertical oscillation frequency in the disc
and dynamical equilibrium cannot be achieved in this short time.
In Fig.\ \ref{figfw2} we show for model A that the
distribution functions $f(|W|)$ including the turnoff stars agree just as well 
with the theoretical distribution functions using the appropriate lifetimes. 

\begin{figure}
\centerline{\resizebox{0.98\hsize}{!}{\includegraphics[angle=0]{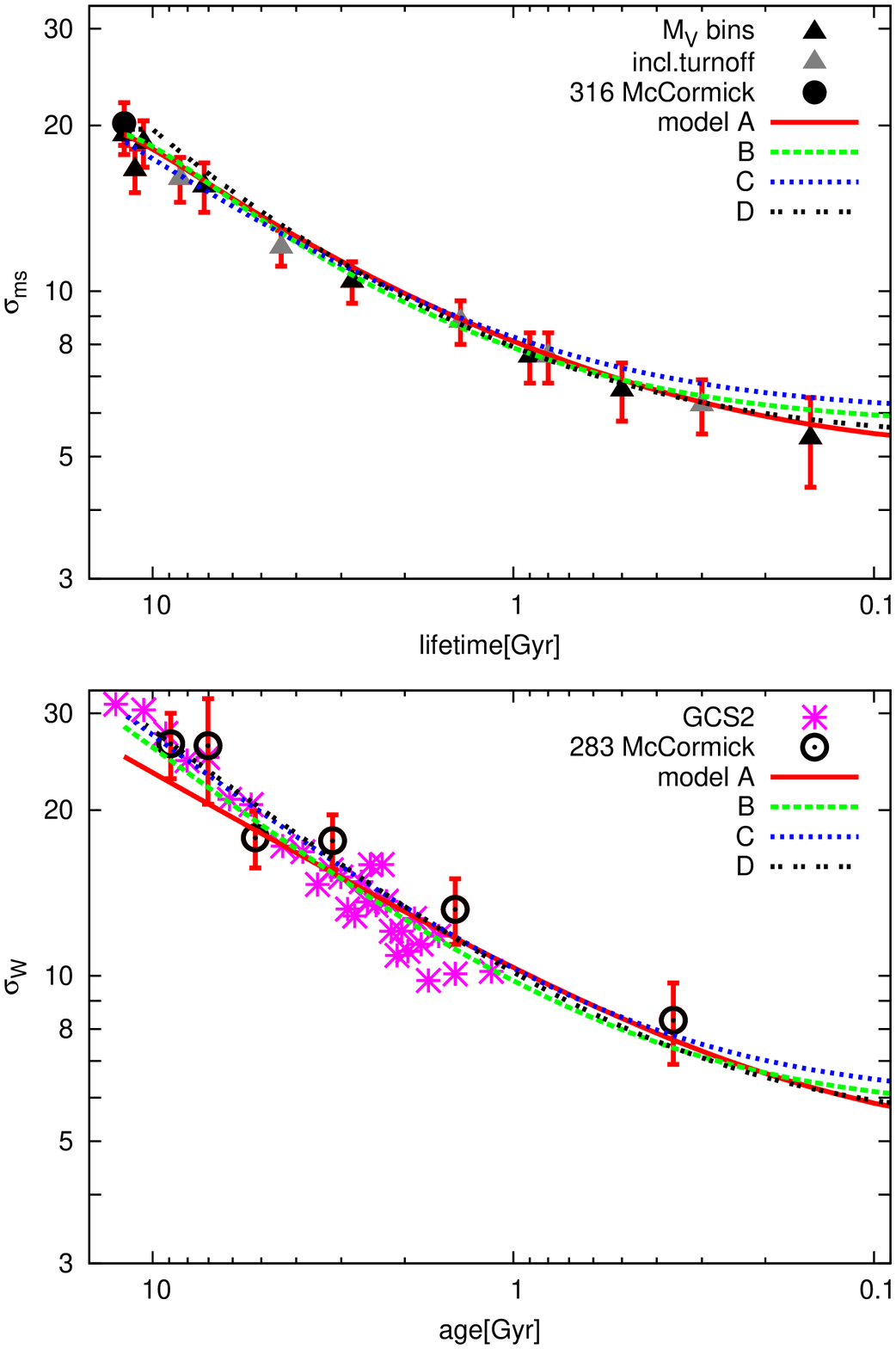}}}
\caption[]{Upper panel: The symbols show the velocity dispersion of the stars along the
main sequence (magnitude bins) as a function of mean lifetime determined from
stellar evolution (Fig.\ \ref{figvage}). The full line
gives the result of the fiducial model A. 
The other lines are for the comparison
models B--D.
Lower panel: The lines show the velocity dispersion as a function of age (AVR) 
for the stellar sub-populations of the models. Full symbols are the age groups of the McCormick K and M dwarfs
\citep{jah97} and the
asterisks are from GCS2.
}
\label{figsig}
\end{figure}

The AVR can be observed only by kinematically unbiased stellar sub-samples with
direct age determinations. In the lower panel of figure \ref{figsig} 
the AVRs of the models A--D are reproduced from figure \ref{figsfr} in
log-scale and compared to two data sets. The circles are the McCormick K
and M dwarfs with ages determined from H and K
line strength \citep{jah97} and the asterisks represent the 
F and G stars of GCS2 with good age determinations.
The ages of the McCormick stars were derived by adopting a constant SFR and
should therefore be consistent with model C. Corrections for a
declining SFR as in the other models would shift the mean ages of the bins
systematically to higher values.
It is interesting to note that the best AVR for models B and C are the same. 
The best fit scaling $\sigma_\mathrm{e}$ is different. This shows
that the variation of the SFR has only a small influence on the shape of the AVR. 
The velocity dispersions of the sub-samples of MS stars with no age
subdivision are determined by the weighted mean over the lifetime with the local
age distribution.

For the AVRs of the different models we find power laws with 
indices $0.375<\alpha<0.5$.
This is in the range of the classical
value of 0.5 \citep{wie77}, of 0.53 \citep{hol09}, and of 0.45 
derived by \citet{aum09}. The best fit slope depends strongly on the zero
point for newly born stars. A higher initial velocity dispersion
$\sigma_\mathrm{W,0}$ leads to a
shallower heating function described by a larger power law index. 
All our models require a large 
$\sigma_\mathrm{W,0}$ and a steep rise of the AVR to reproduce the distribution
functions of stars with lifetimes 0.5-3\,Gyr. The maximum velocity dispersion
of the oldest thin disc stars is 25-30\,km/s dependent on the SFR.

 The upper panel of Fig.\ \ref{figsig} shows the excellent 
agreement of the model with the data from the nearby stars. 
 The black triangles are the velocity dispersions in the magnitude bins
 of the main sequence stars with lifetimes from Fig.\  \ref{figvage}. The full
 circle is the velocity dispersion of the McCormick stars. The underlying
 distribution functions of these data sets were used for the best fitting.
 The grey triangles are the velocity dispersions including the turnoff stars in
 the five magnitude bins with $M_V<4.5$\,mag. As expected, they show a 
 systematically larger
 velocity dispersion and have larger lifetimes compared to the pure MS samples.
 These data confirm the underlying assumption of a continuous disc heating and
 dynamical equilibrium of the stellar sub-samples.

\begin{figure}
\centerline{\resizebox{0.98\hsize}{!}{\includegraphics[angle=0]{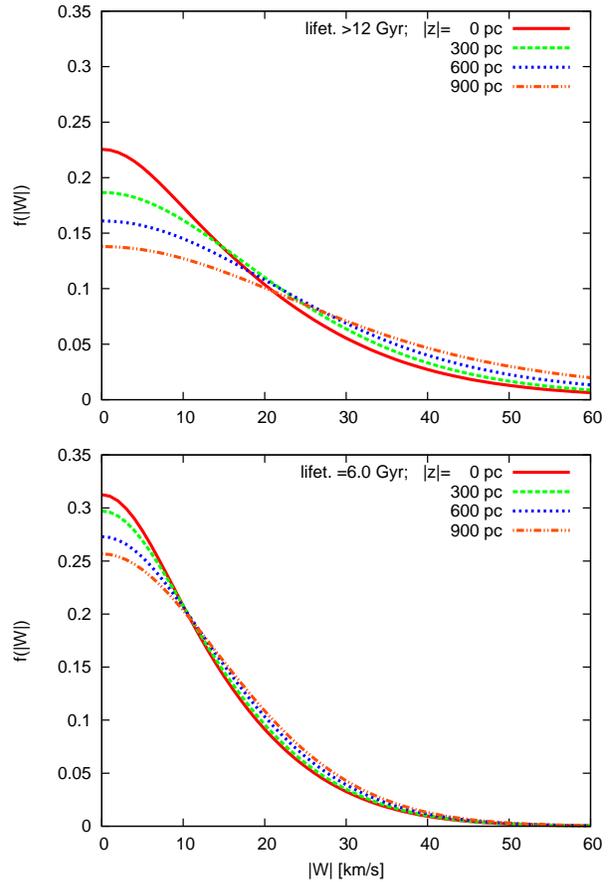}}}
\caption[]{The normalised velocity distribution functions
$f(|W|)$ (for $dw=5$\,km/s) at different heights $z$ for model A. 
The upper panel shows of stars with lifetimes larger than 12\,Gyr
 including the thick disc contribution, whereas the lower panel is 
for thin disc stars with lifetime 6.0\,Gyr.}
\label{figfwz}
\end{figure}

Since the shape of $f(|W|)$ depends on the age distribution of the stars, it depends also on
the height $z$ above the mid-plane. Fig. \ref{figfwz} shows the
variation above the plane for stars with lifetimes larger than 12\,Gyr
(upper panel) and stars with a lifetime of 6\,Gyr
(lower panel; corresponding to
magnitudes $M_\mathrm{V}=3.5- 4$\,mag).
The vertical gradient in the velocity dispersion of all thin disc stars is
slightly stronger than that of the stars with lifetimes larger than 12\,Gyr,
because the dip near the mid-plane is enhanced due to the contribution of stars
with smaller lifetimes. The profile is shown in figure \ref{figrhoz} for models
A and C. The large fraction of young stars in model C with constant SFR leads to
a significantly larger gradient.

\subsection{Density profiles \label{density}}

\begin{figure}
\centerline{\resizebox{0.98\hsize}{!}{\includegraphics[angle=0]{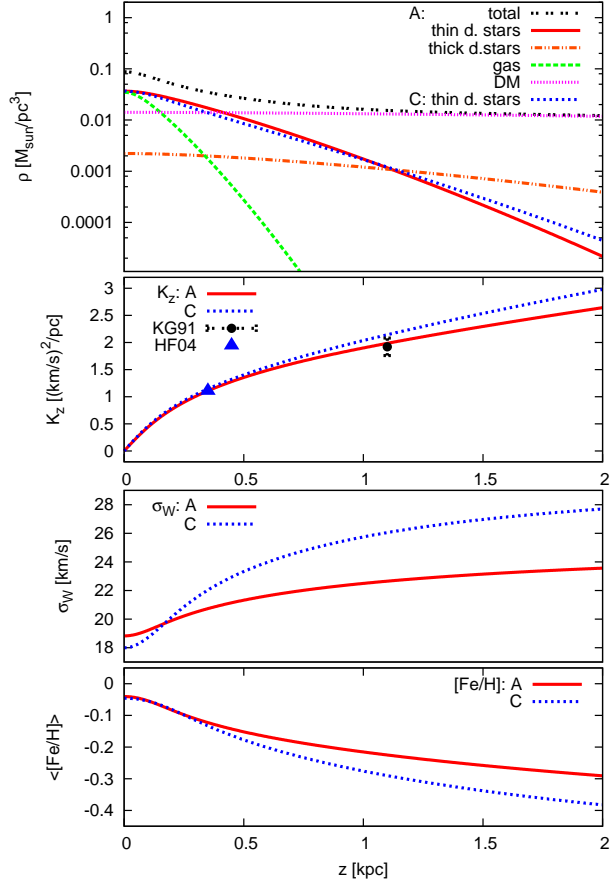}}}
\caption[]{Vertical profiles of the model A and of the thin disc of 
model C (SFR=const.). 
From top to bottom: 
Density profiles; 
$K_\mathrm{z}$ force complemented by the corresponding surface densities of
\citet{kui91} and \citet{hol04};
velocity dispersion of the thin stellar disc;
V-band luminosity weighted mean metallicity of the thin stellar disc.
}
\label{figrhoz}
\end{figure}

The shape of the vertical density profiles of all components and sub-populations
are determined by their kinematics and the self-consistent gravitational
potential. A general feature of the vertical density profiles of the stars and of 
the gas is the flattening at the
galactic plane. All profiles are between an exponential profile and that of 
an isolated isothermal profile given by a sech$^2$ function. In the upper panel
of figure \ref{figrhoz} the profiles of model A are shown. The stellar disc
profile of model C (also plotted in figure \ref{figrhoz}) is significantly
flatter and deviates from model A in the regime $|z|\sim 500$\,pc and above  
$|z|\sim 1500$\,pc. The second plot in figure \ref{figrhoz} shows the difference
in the $K_\mathrm{z}$ force that is a measure of the total surface density up to
$|z|$, of models A and C. Below 500\,pc the profiles are very similar despite
the differences in the mass fractions and the kinematics. The corresponding
surface density values of \citet{kui91} and \citet{hol04} are added.

For the gas, the thickness is
$\sim$150\,pc compared to the exponential scale height of $\sim$100\,pc. 
The density profile and 
the surface density are consistent with the observed HI-profile 
\citep{dic90}: 0.014\,$\msun/\mathrm{pc}^3$, 
5.0\,$\msun/\mathrm{pc}^2$, 177\,pc for the mid-plane density, surface density,
thickness) adding about 50\% of H$_2$ with smaller scale height
\citep{bro88} and applying the correction factor of 1.4 for
Helium and heavy elements.

Due to the gravitational potential of the disc, the local density of the dark matter halo is
$\sim$50\% larger than the DM density at $z_\mathrm{max}$ and $\sim$20\% larger
than the mean DM density given by $\Sigma_\mathrm{h}/(2z_\mathrm{max})$.

\begin{figure}
\centerline{\resizebox{0.98\hsize}{!}{\includegraphics[angle=0]{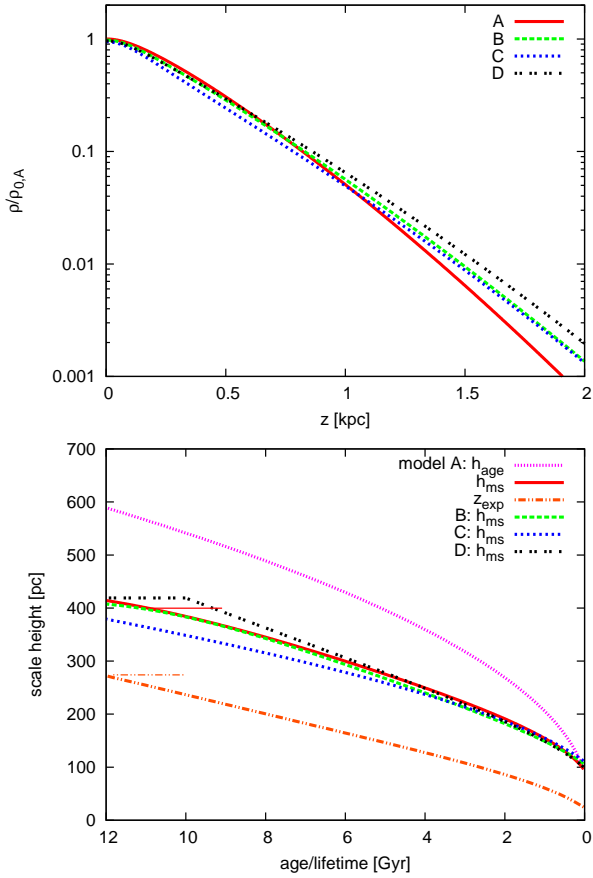}}}
\caption[]{Top panel: Comparison of the density profiles for thin disc stars with lifetimes 
larger than the age of the disc. 
Bottom panel: The (half-)thickness of the sub-populations as a function
of age $h_\mathrm{age}$ and of the MS stars as a function of lifetime
 $h_\mathrm{ms}$. For comparison the exponential scale heights $z_\mathrm{exp}$
 for the MS stars are included. The horizontal lines are
  the overall thickness $h_\mathrm{eff}$ 
  and the exponential scale height $z_\mathrm{s}$ of all stars.}
\label{figrhohz}
\end{figure}

A measure of the profile flattening is given by the ratio of the
thickness and the exponential scale height. The half-thickness of MS stars
$h_\mathrm{ms}$ as function of lifetime is shown in  the lower panel
of Fig. \ref{figrhohz} for all models A--D. The differences above a lifetime of
6\,Gyr is significant leading to differences in the determination of surface
densities from local densities as used in the estimation of the SFR and IMF from
local data. For model A additionally the exponential scale height $z_\mathrm{exp}$
as function of lifetime and the half-thickness $h_\mathrm{age}$ as function
of age are plotted. For young populations
$z_\mathrm{exp}$ is much smaller than $h_\mathrm{ms}$, because the density
profiles have Gaussian and not exponential wings. For longer
lifetimes $h_\mathrm{ms}$ is $\sim$50\% larger than $z_\mathrm{exp}$.

\begin{figure}
\centerline{\resizebox{0.98\hsize}{!}{\includegraphics[angle=0]{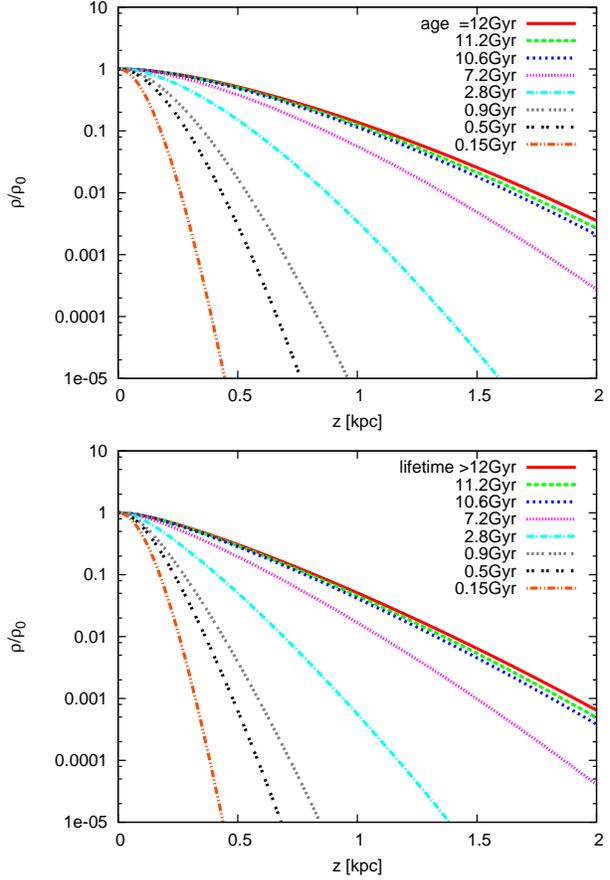}}}
\caption[]{Upper panel: Normalised density profiles of the stellar
sub-populations as a function of age.
Lower panel: Same for the main sequence stars as a function of lifetime. The set
of lifetimes corresponds to the mean lifetimes of the sub-samples used for the
velocity distribution functions (see Fig.\ \ref{figfw}).}
\label{figrho}
\end{figure}
The density profiles of MS stars differ in shape from the profiles of
the sub-populations of single ages. The lower panel of Fig.\ \ref{figrho} shows
the normalised density profiles with lifetimes according to the samples used for
 the model. They are
steeper than the corresponding profiles of the sub-populations with the same age
and significantly shallower than the density profile using the mean age of the
sub-population. This difference is quantified by the thicknesses $h_\mathrm{age}$ and
$h_\mathrm{ms}$ shown in the lower panel of Fig. \ref{figrhohz}.

\begin{figure}
\centerline{\resizebox{0.98\hsize}{!}{\includegraphics[angle=270]{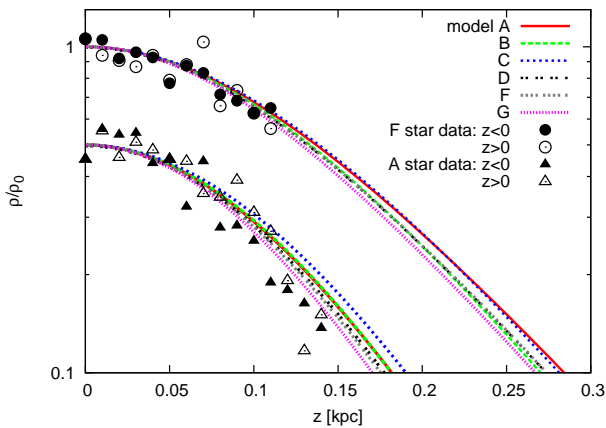}}}
\caption[]{Density profiles for the A and F star sample from \citet{hol00} with
an arbitrary vertical offset (full and open symbols below and above the
mid-plane, respectively. The lines are models A--D and F,G for corresponding
mean lifetimes 0.4\,Gyr and 1.0\,Gyr for the A and F stars, respectively. 
The profiles for the A stars are shifted
additionally by a factor of two.
}
\label{figAF}
\end{figure}

An independent test of the model is the comparison with directly observed density
profiles of MS stars as derived by
\citet{hol00} for two sub-samples of A stars (with 
$0<M_{V}<1$\,mag) and early F stars (with $1<M_{V}<2.5$\,mag). 
The mean stellar lifetime
in these magnitude bins are 0.3\,Gyr and 1.0\,Gyr, respectively, 
including turnoff stars. 
In order to minimise the systematic asymmetry of the
profiles (mainly for the A stars) we applied an additional offset of the solar
position in $z$ of +5\,pc. 
The result is shown in figure \ref{figAF}. For the A stars all profiles are
shifted by a factor of two to avoid overlaps. Over-plotted are the normalised
profiles of models A--D and F,G.
The observed profile for the F stars is in very good
agreement with all models. The observed profile of the A stars 
is slightly steeper than the model and requires a larger local density
or an even shorter lifetime. Models F and G show the effect of a larger local
density in two steps (see table \ref{tabmod}). Since we do not expect that these
stars are already in dynamical equilibrium, the deviations are in a reasonable
range.
In the Besan\c{c}on model \citep{rob03} 
Einasto laws are used for the density distribution of isothermal sub-populations.
The thicknesses (=flattening in their model) are determined by solving the
Poisson equation. 
The shapes of the density profiles deviate systematically from our results and
this means that dynamical equilibrium is achieved only approximately.
A detailed investigation of
the differences to our model is beyond the scope of this paper.

\subsection{Age distributions \label{ages}}

\begin{figure}
\centerline{\resizebox{0.98\hsize}{!}{\includegraphics[angle=0]{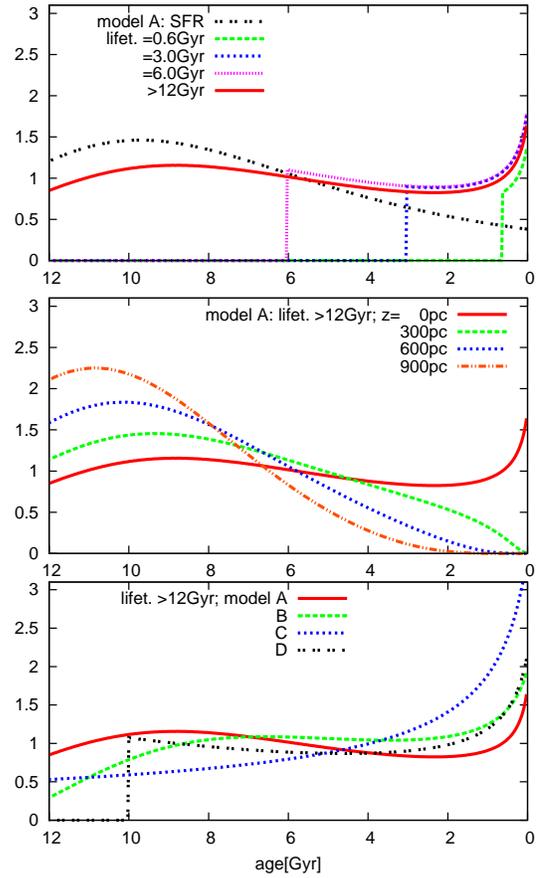}}}
\caption[]{The upper panel shows the normalised local age distributions of 
MS stars with different lifetimes.
The middle panel shows the age distribution of stars with lifetime larger than
12\,Gyr at different heights above the Galactic plane.
The lower panel shows the 
local age distributions of models A--D.}
\label{figageh}
\end{figure}

The age distributions of the stellar sub-samples of MS stars depend on the
lifetime and on the vertical structure. The upper panel of figure \ref{figageh}
shows for model A the normalised local age distribution for different lifetimes.
The comparison with the SFR shows clearly the over-representation 
of young stars in the solar  neighbourhood due to the 
increasing thickness with age.
The vertical dilution is quantified by the (half-)thickness $h_\mathrm{ms}$
(lower panel of Fig.\ \ref{figrhohz}).

A comparison of the local age distributions of stars with lifetime
larger than the age of the disc for models A--D are shown in
the lower panel of figure \ref{figageh}.
In model A the local age distribution of stars with lifetime larger
than the disc age varies by
less than a factor of two around the mean value. \citet{bin00}
and \citet{cig06} propose a constant local age distribution in
this sense. This is consistent with our model 
taking the uncertainties of isochrone ages into account. 
In model B there is a lack of old stars which cannot be tested by direct
measurements,
because for this age range the age determinations are very uncertain.
Model C with constant SFR predicts a
strong dominance of young stars in contrast to the findings of \citet{bin00}
and \citet{cig06}.

The age distributions are a strong function of $z$ above the plane. The middle
panel of Fig.\ \ref{figageh} shows the lack of young stars in steps of $\Delta
z=300$\,pc above the mid-plane.

\subsection{Metallicity \label{feh}}

\begin{figure}
\centerline{\resizebox{0.98\hsize}{!}{\includegraphics[angle=0]{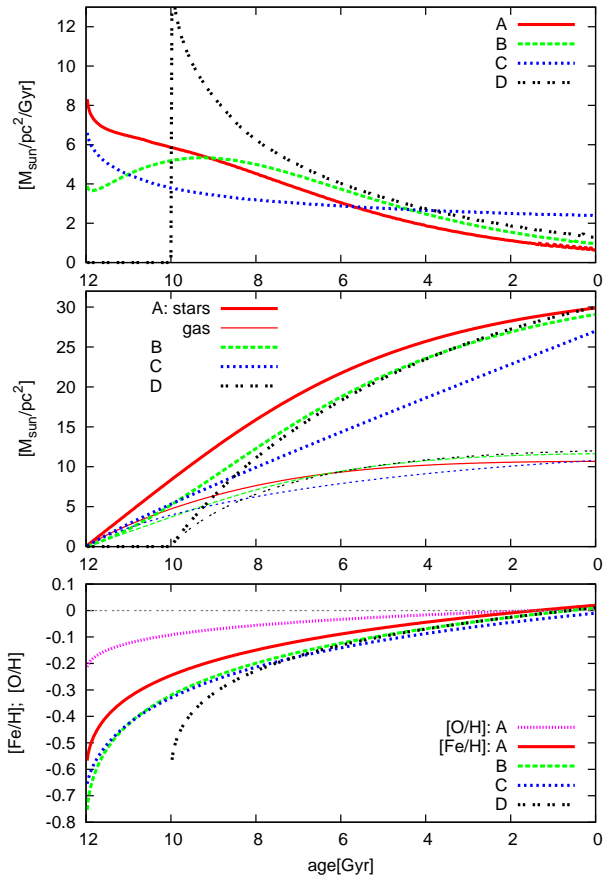}}}
\caption[]{ 
Upper panel: Gas infall rates for models A--D.
Middle panel: Cumulative stellar and gas surface density (thick and thin
lines, respectively).
Lower panel: Metal enrichment laws [Fe/H] of models A--D. 
The horizontal
thin black line at solar metallicity is for guiding the eye.
Additionally the oxygen enrichment [O/H]=0.375[Fe/H] is shown for model A
(pink dotted line).
}
\label{figevolv}
\end{figure}

Since MS luminosities and lifetimes depend on metallicity, we include a simple
analytic metal enrichment law $\mathrm{[Fe/H]}(t)$. 
We adopt a generalised form of a closed box model with a Schmidt star 
formation law for the oxygen abundance [O/H]$(t)$ \citep{jus96} and
transform it by a linear approximation to [Fe/H] from \citet{red03} for thin
disc stars, i.e.
\bqn
\mathrm{[Fe/H]}(t)&=&2.67\mathrm{[O/H]}(t) \nonumber\\
\mathrm{[O/H](t)}&=&\lg \left(Z_\mathrm{oxy}(t)\right)\\
Z_\mathrm{oxy}(t)&=&Z_\mathrm{oxy,0}+\left(Z_\mathrm{oxy,p}-Z_\mathrm{oxy,p}\right)
	\frac{\lg \left[1+\ln(1+q\,(t/t_\mathrm{p})^r\right]}{\ln(1+q)}\nonumber
\eqn
There are four fitting parameters 
$q,r,\mathrm{[Fe/H]}_0,\mathrm{[Fe/H]}_\mathrm{p}$ which determine the scaling
and the curvature. 
We tested a large range of parameters: 
initial metallicity $\mathrm{[Fe/H]}_0$=-0.6\dots -0.8,
present day metallicity $\mathrm{[Fe/H]}_\mathrm{p}$=-0.1\dots +0.2,
linear increase up to steep early enrichment covering 
$\mathrm{[Fe/H]}$=-0.4\dots +0.1 at an age of 6\,Gyr. In order to reproduce the
high metallicity tail without intrinsic scatter we investigated also some
AMRs with an additional accelerated enrichment in the last 2 Gyr. 
In every case it turned out
that the local G dwarf metallicity distribution is a strong restriction for the
AMR for any given SFR and AVR.
The parameters for models A--D are
\bqn
A&:& q=2 \, ; \,  r=0.55\, ; \, \mathrm{[Fe/H]}_0 =-0.6\, ; 
\,\mathrm{[Fe/H]}_\mathrm{p} =0.02\nonumber \\
B&:& q=2 \, ; \,  r=0.55\, ; \, \mathrm{[Fe/H]}_0 =-0.8\, ; 
\,\mathrm{[Fe/H]}_\mathrm{p} =0.01\nonumber \\
C&:& q=3 \, ; \,  r=0.70\, ; \, \mathrm{[Fe/H]}_0 =-0.67\, ; 
\,\mathrm{[Fe/H]}_\mathrm{p} =-0.01\nonumber \\
D&:& q=2 \, ; \,  r=0.55\, ; \, \mathrm{[Fe/H]}_0 =-0.6\, ; 
\,\mathrm{[Fe/H]}_\mathrm{p} =0.01\nonumber 
\eqn
The metal enrichment laws are shown in the lower panel of 
figure \ref{figevolv}. For model A [O/H] is also plotted.
The local metallicity distribution is very sensitive to the
present day metallicity due to the shallow slope. 

\begin{figure}
\centerline{\resizebox{0.98\hsize}{!}{\includegraphics[angle=0]{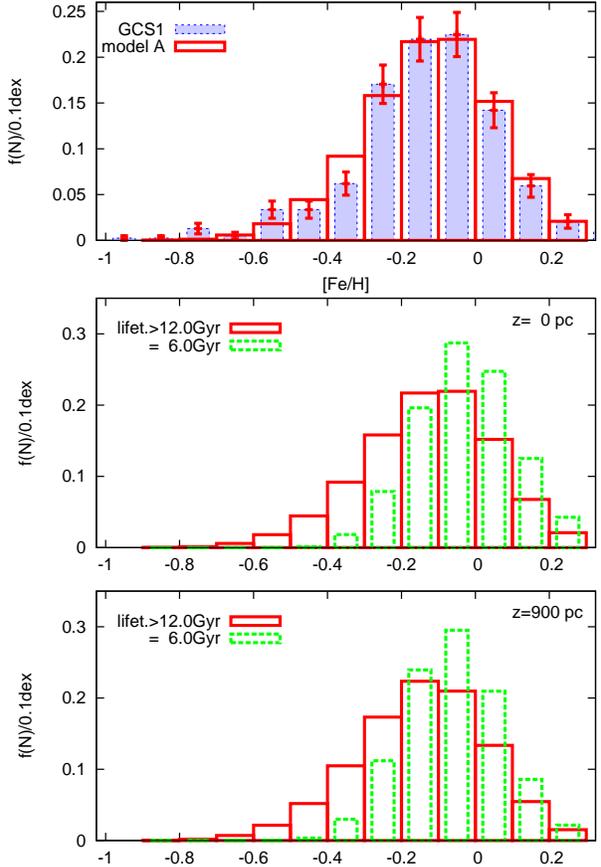}}}
\caption[]{Upper panel: The metallicity distribution of model A compared to
the observations from GCS1
in the stellar mass range $0.84\le M/\msun\le 0.90$.
Lower panels: The metallicity distribution for dwarfs with lifetime 6.0\,Gyr
in comparison to the dwarfs with lifetime $>12$\,Gyr in the
solar neighbourhood and 900\,pc above the mid-plane.}
\label{figfeh}
\end{figure}

From the metal enrichment the local
metallicity distribution for stars with lifetime larger than the age of the
disc is calculated. Before binning the theoretical distribution
we add a Gaussian scatter
which represents an intrinsic scatter and observational errors. 
We tested a range up to 0.2\,dex and found a best value of 0.13\,dex
corresponding to a FWHM=0.31\,dex in order to reproduce both the narrow width
of the maximum and the high metallicity wing in the observations
(see figure \ref{figfeh}).
The local metallicity distribution is determined from the
Copenhagen F and G  star sample \citep{nor04} selecting all
stars with masses $0.84\le M/\msun\le 0.90$ up to the completeness limit
$r<40$\,pc. 
The lower mass limit of $0.84\msun$ was chosen in order to avoid incompleteness
for the less luminous metal rich stars. The upper limit was set to be
consistent with a lifetime larger than 12\,Gyr for the selected stars.

For model A the comparison of the derived local metallicity distribution for stars
with adopted lifetime larger than 12\,Gyr and the data of the selected mass bin 
is shown in the upper panel of Fig.\ \ref{figfeh}. The results for models B--D
are very similar.
The middle panel of Fig.\ \ref{figfeh} shows the predicted metallicity 
distribution in model A for stars with lifetime 6\,Gyr compared to the observed
distribution with lifetime  larger than 12\,Gyr in the solar neighbourhood.
The lower panel shows the same comparison but at $|z|= 900$\,pc.

The metal enrichment [O/H] and the SFR can be embedded in a local chemical 
evolution model with primordial gas infall. 
Since the gas infall rate is an additional free function, the metal enrichment
cannot be derived directly from the SFR, IMF and stellar yields.
We proceed in the following way. For the oxygen
enrichment we adopt instantaneous recycling and mixing to determine the infall
rate of primordial gas. The mass loss from stellar evolution is taken into
account.  
Then the oxygen yield in solar units is given by 
\bq
Y_\mathrm{oxy}=\langle Z_\mathrm{oxy}\rangle +Z_\mathrm{oxy,p}
\frac{\Sigma_\mathrm{g}}{S_0}+Z_\mathrm{oxy,0}\frac{G_0}{S_0}
\eq
Here $\langle Z_\mathrm{oxy}\rangle$ is the mean metallicity of all born stars
(not in log-scale and weighted by mass), $G_0$ is the initial surface density
of gas, $\Sigma_\mathrm{g}$ the present day value
and $S_0$ the total amount of born stars.
We start with a negligible initial amount of gas $G_0$ and find for the yields
in solar units of models A--D $y_\mathrm{oxy}=1.1135, 1.1301, 1.1290, 1.1350$, respectively.
The infall rate, the surface density
of gas and stars and the metal enrichment laws are shown in Fig.\ \ref{figevolv}. 

\subsection{Thick disc \label{thd}}

The thick disc properties in the solar neighbourhood are not well determined. 
In many investigations a local stellar contribution of $\sim 5\%$ and a velocity
dispersion of $\sim$45\,km/s are adopted. These values lead to a dominance of
thick disc stars compared to the thin disc at distances $z>1$\,kpc which is
directly observed for K dwarfs \citep{phl05}. But in a
recent paper local contributions up to 20\% and metallicities  up to solar
are claimed \citep{ben07}. Since the method of the latter paper relies on a
statistical separation from the kinematics by adopting Gaussian velocity
distribution functions for the thin and for the thick disc, it underestimates strongly the 
thin disc contribution at the high velocity tail.
In our models we use standard values for the thick disc and investigate the
influence of the thick disc on the other model properties. 
We split the local stellar density into a thin disc and a thick disc
component. For models A--D we
set $\sigma_\mathrm{t}\sim 45$\,km/s and choose
$\Sigma_\mathrm{t}$ leading to a local stellar mass contribution of $\sim5\%$.
The local mass fraction of the thick disc equals the
thin disc density at $z= 1.1-1.25$\,kpc in the models. We adopt an age of
12\,Gyr for the thick disc leading to two consequences. Firstly the high mass
end of the MS is missing and the fraction in number of low mass stars is a
factor of 1.84 larger than the mass fraction. This has to be taken into account
when comparing to results of star counts. For model A the number of K dwarfs
in the thick disc equals that in the thin disc at $z=950$\,pc very close to 
the observed K dwarf density profile of \citet{phl05}. Secondly for the
kinematics $f(|W|)$ the thick disc
contribution is only taken into account for stars with lifetime larger than
12\,Gyr. Here the
correction factor 1.84 is also included.

In all cases the density profile of the thick disc can be fitted by a
sech$^\alpha(z/\alpha h_\mathrm{t})$ profile to better than 3\%. The
corresponding best fit coefficients are given in table \ref{tabmod}. For the
thin disc profile a corresponding fit deviates up  to 25\%.

\begin{figure}
\centerline{\resizebox{0.98\hsize}{!}{\includegraphics[angle=0]{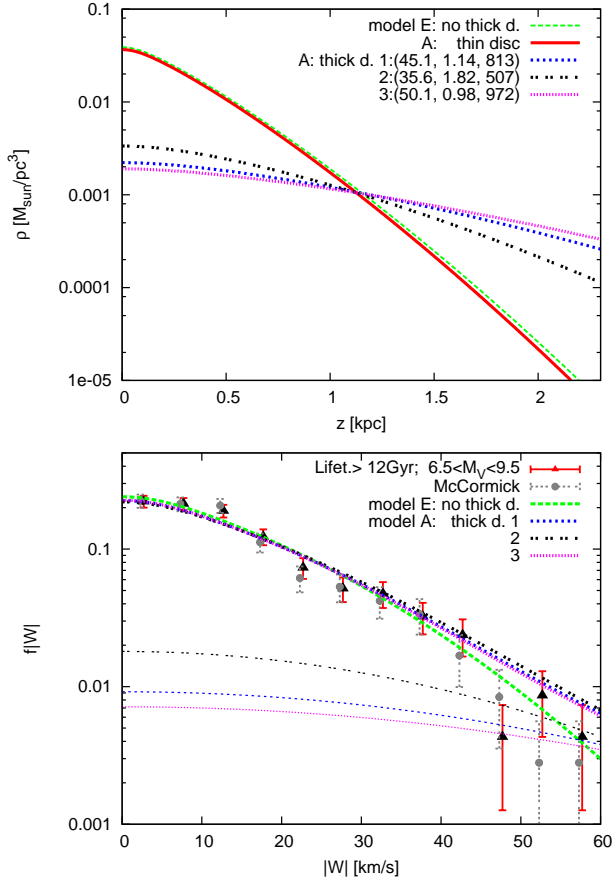}}}
\caption[]{Upper panel: Density profiles of models with different isothermal thick disc 
components:  Thin disc stars of model E without thick disc and  model A with
different thick discs;
 thick disc parameters
($\sigma_\mathrm{t},\alpha,h_\mathrm{t}$) are indicated.
Lower panel: 
Reproduction of $f|W|$  from the last two plots of figure \ref{figfw}, but
in log-scale to demonstrate the effect of the thick disc contribution. Thick
lines are the combined distribution functions and thin lines are the thick discs
only.
 }
\label{figthd}
\end{figure}

 We have calculated two alternative thick discs in model A with smaller and larger velocity dispersion,
 respectively. The surface density of the thick discs are adjusted to yield the
 same crossing point of the thin and thick disc profile (upper panel of figure
\ref{figthd}). the local density contributions of the thick discs are 5.7\%,
8.6\% and 4.9\% for discs 1,2, and 3, respectively.
The contribution of the thick disc to the velocity distribution function is
shown in the lower panel of Fig.\ \ref{figthd}. All these models are still
consistent with the data in the $|W|$-range of 40--60\,km/s. From the distribution
functions we can conclude that a kinematically distinct thick disc cannot exceed
a local density contribution of 10\% in mass density.

Model E is a model without thick disc component but with the same AVR and SFR 
as in model A.
For the minimum $\chi^2$ no alteration in the AVR and SFR are needed. The best
fit $\sigma_\mathrm{e}$ is slightly higher. The total star formation $S_0$ and
the densities of the thin disc are larger containing now all stars in the solar
neighbourhood.
The corrections to all other parameters are within a few percent.
The effect of the thick disc on the velocity
distribution function of stars with lifetime larger than 12\,Gyr is shown in the
lower panel of Fig.\ \ref{figthd}. Only the high velocity tail
is affected.
 
\subsection{Initial mass function \label{imf}}

\begin{figure}
\centerline{\resizebox{0.98\hsize}{!}{\includegraphics[angle=0]{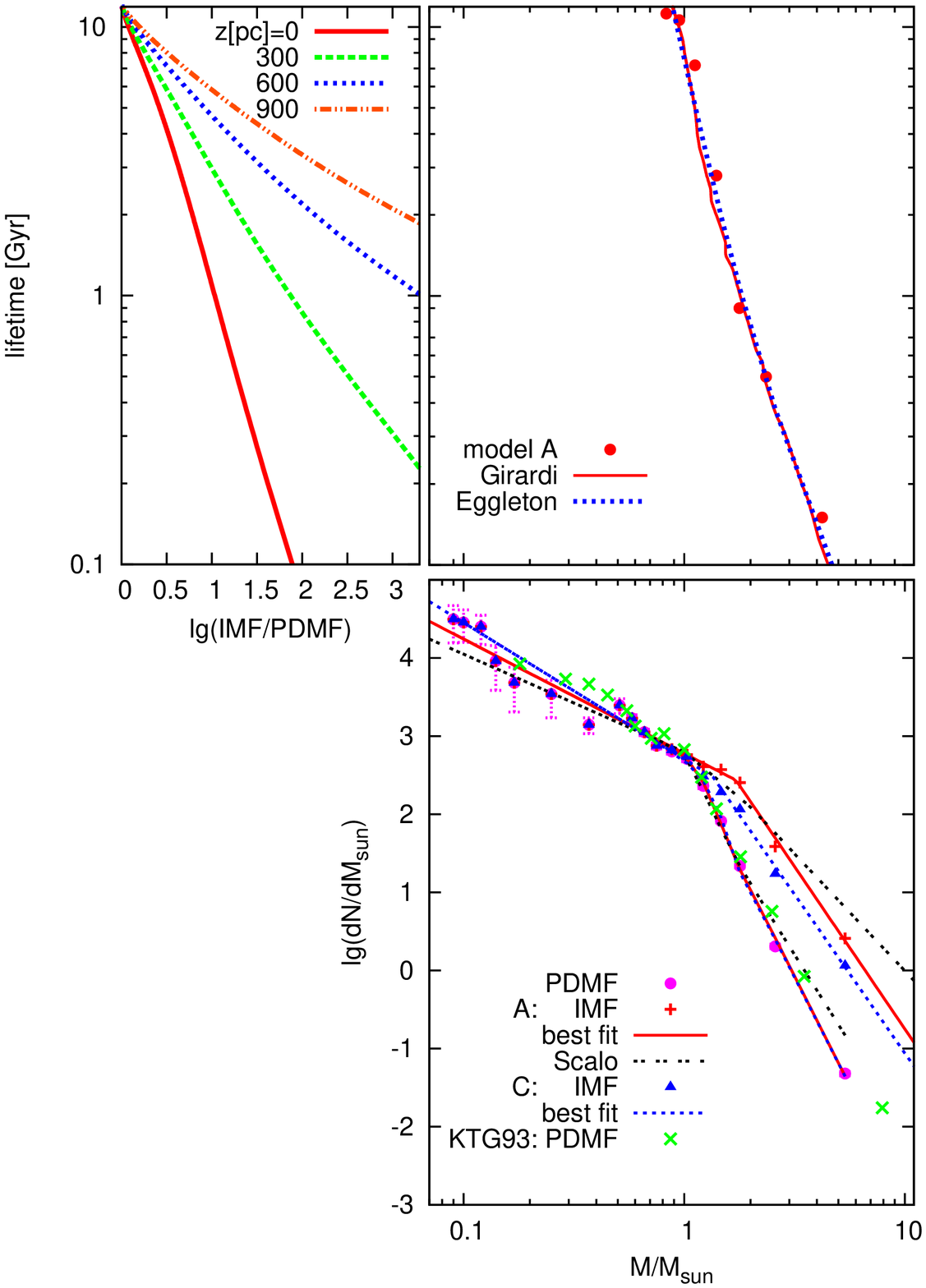}}}
\caption[]{ 
Upper left panel: Conversion factor (=IMF/PDMF) for model A that must be applied to 
the PDMF to derive the normalised IMF from local star counts.
Upper right panel: Stellar lifetimes as function of stellar mass to relate the
conversion factors at the left to the PDMF below. Dots are the values used in
the models (from Fig. \ref{figvage}) and lines are
values determined by \citet{gir04} and \citet{egg89}.
Lower panel: Full circles show the PDMF of our sample and 
crosses the PDMF of KTG93. 
The IMF from models A and C are given by the plus signs and the triangles,
respectively.
The best fit IMF and corresponding PDMF for model A
(full red lines) and model C (dotted blue lines) are over-plotted.
 The dashed black lines show the Scalo IMF and IMF for comparison.
}
\label{figpdmf}
\end{figure}

We have fixed the IMF for the best fitting process (equation \ref{eqscalo}).
We can use the local luminosity function to test the adopted IMF.
For the determination of the IMF from the local luminosity function, the
luminosity function is first converted to a mass function $dN=f(M)dM$ using the
transformation formulae of \citet{hen93}, corrected in 
\citet{hen99}, and extended to bright stars according to 
Schmidt-Kaler in \citet{lan82}, which was confirmed by \citet{and91}. 
The star numbers are normalised to a sphere with radius
R=20\,pc. The result is shown in the lower panel of
Fig. \ref{figpdmf} and compared to the PDMF given in \citet{kro93} (KTG93). 
In our model we use the system mass function excluding resolved B components of
binaries. In addition turnoff stars were excluded to be consistent with 
the main sequence lifetimes used in the dynamical model. 
Therefore our PDMF is systematically below KTG93. The PDMF is also corrected
for the vertical gradient of the density profiles in the observed volume.
In the brightest bin with lifetime 0.15\,Gyr the density in the mid-plane is
50\% larger than the mean density in the 200\,pc sphere. 

In order to determine the conversion factor from the local PDMF
to the IMF due
to the finite lifetime and the vertical thickness $h_\mathrm{ms}$, the main
sequence lifetime is needed. We use the lifetimes estimated from the
 stellar evolution tracks shown in Fig. \ref{figvage} for the different 
magnitude bins, where
the metal enrichment and the relative weighting due to the increasing number of
stars with decreasing mass in the mass interval is taken into account. These
data are shown in the upper right panel of Fig. \ref{figpdmf} with a
comparison of the lifetimes determined directly from evolutionary tracks of 
\citet{gir04} and with the analytic fitting formula of \citet{egg89}. 
Any systematic variation of the lifetimes result in
significant changes of the conversion factor and therefore the IMF.
The conversion factors are a combination of the
dilution by the increasing thickness with age (measured by $h_\mathrm{ms}$) and
the fraction of born stars still on the main sequence due to the finite
lifetime. The last factor depends strongly on the SFR. The upper left panel of 
Fig. \ref{figpdmf} shows the conversion factor (=IMF/PDMF) as a 
function of lifetime at different heights above the mid-plane for model A.

In the lower panel of Fig. \ref{figpdmf} the PDMF and the IMFs of model A and
C are shown with different symbols. 
The IMFs of model B and D are very similar to that of model A.
We compared different best fits in
log-scale to test the significance of different IMFs.
The highest mass bin is not consistent with the Scalo IMF (Eq. \ref{eqscalo}),
but the dominating contribution to the rms value of 4.98 originates from the
mass range $0.4<M/\msun<1$. A similar best fit of a Kroupa IMF KTG93 
\citep{kro93} leads to a
rms value of 5.67 for $\alpha=-2.7$ at  $M/\msun>1.0$. 
If we allow the slope of the Scalo IMF to vary at the
high mass end for $M/\msun>2.0$ we find a best fit value of 
$\alpha=-4.16$ and a rms of 2.81. 
A similar best fit of a Kroupa-like IMF as used in \citet{aum09} leads to 
$\alpha=-3.06$ at $M/\msun>1.0$ with rms=5.96. 

An inspection of the shape of the IMFs of models A and C shows that
the break near 1\,$\msun$ seems artificial. The break is obvious in the PDMF,
but the reason for this is that the correction factor due to the finite 
lifetime starts to apply above $\sim 1\msun$.
Therefore we determined an alternative IMF by fitting power laws in two mass
 regimes only, where we fixed the break mass $M_0$ by minimising the rms
of the best fit. 
The best fits are
\bqn
\dd N/\msun &=& N_0\left(M/\msun\right)^{\alpha}
	\dd \left(M/\msun\right) \quad\mathrm{with}\\
A:\,\alpha&=&\left\{\begin{array}{rcl}
        -1.46\pm 0.10&&0.08\le M/\msun < 1.72\\
        -4.16\pm 0.12&&1.72\le M/\msun < 6.0
	\end{array}\right.\nonumber\\
&&N_0=272/\msun\quad \mathrm{at}\quad M_0/\msun=1.72\,,\nonumber\\
C:\,\alpha&=&\left\{\begin{array}{rcl}
        -1.70\pm 0.13&&0.08\le M/\msun < 1.35\\
        -4.07\pm 0.11&&1.35\le M/\msun < 6.0
        \end{array}\right.\nonumber\\
&&N_0=388/\msun\quad \mathrm{at}\quad M_0/\msun=1.35\,,\nonumber
\eqn
where $N_0$ gives the normalisation in the 20\,pc sphere. The slopes at the
high mass end are very similar for models A--D, because the IMFs are
essentially shifted vertically relative to each other. The best fits of the
Kroupa-like IMFs show that forcing the slope in the mass range $1-1.5\msun$
to be the same as at the high mass end results in a flatter IMF and a much
larger rms value.
At that stage of the investigation we abstain from final conclusions, because
we did not discuss possible biases in the PDMF from local star counts.
We expect that the feedback of a corrected IMF with a steeper slope at the high mass end
to the disc model via the mass loss is
very small. Additionally it affects only the velocity distribution functions
for lifetimes smaller than $10^8$\,yr significantly. 

The strongest constraints on the IMF at high masses and the present day SFR
is the observed number of A and late B stars in the solar neighbourhood. Since the bright
stars with $M_\mathrm{V}<0.5$\,mag are observed in a sphere with a radius of
200\,pc, the sample size is a direct measure of the local surface density. Therefore
the conversion from the IMF to the mean SFR in the last few 100\,Myr 
depends only on the lifetime of the stars. That means, a higher present day
SFR requires a shorter lifetime for the A stars or a steeper IMF.

As an example the constant SFR used in the Besan\c{c}on model \citep{rob03} 
relies mainly on the assumption of a steep IMF in the stellar mass range 
1-3\,$\msun$ \citep{hay97b}. 
This model and also the SFR determination of 
\citet{ver02} show that the slope of the IMF in the mass range of
1-3\,$\msun$ that covers roughly the lifetime range of 0.2-10\,Gyr, is crucial
for the derivation of the long-term SFR from star counts. A steep IMF
in that mass range introduces a strong bias to young ages, because the observed number of
stars above 2\,$\msun$ 
predict too many young lower mass stars via the steep IMF. This  effect is
strengthened by the fitting procedure, because direct age
determinations by isochrone fitting are less significant in the lower mass range.
The result is a
bias to young ages in the stellar mass range covering the lifetime above 1\,Gyr.
An underestimation of the SFR for ages above 1\,Gyr is the consequence.

The position of the Sun is probably about 20\,pc above the mid-plane
\citep{hum95}.
For the determination of the mid-plane density this offset can play a
significant role for sub-populations with small scale
height ($M_\mathrm{V}<1.5$\,mag). We corrected for that implicitly, 
since we determined the
mid-plane density of these magnitude bins for the PDMF using spheres with large
radii (see Table \ref{tabdata}), where the offset can be neglected.

\section{Summary} \label{summary}

We presented a new disc model for the thin disc in the solar cylinder
 based on a continuous  star formation history (SFR) 
 and a continuous dynamical heating law (AVR) of
the stellar sub-populations.
This new model combines and improves the advantages of several 
different attempts to model
the vertical structure in the solar neighbourhood:
We used a sequence of isothermal sub-populations in dynamical equilibrium as 
\citet{bah84c} and \citet{aum09} did; 
We used the full velocity distribution functions $f|W|$ as \citet{hol00} did, 
and not only the velocity dispersions (the AVR);
We solved the Poisson equation self-consistently including the thick disc, gas
and DM halo contribution as done in the Besan\c{c}on model \citep{rob03}. 
A chemical evolution model with reasonable gas infall rate, which was tuned 
to reproduce the local [Fe/H] distribution of G dwarfs,
is included. This enables us to apply correct MS luminosities and
lifetimes. Additionally
our model is insensitive to the IMF, because it is based on the normalised
velocity distribution functions of MS stars. 

We determined pairs of (SFR, AVR) by a best fit of the local kinematics.
The SFRs which are consistent with the MS velocity distribution
functions show a decline factor below five down to unity (= const. SFR).
The strongest feature that would distinguish between a constant and a
declining SFR is a direct
determination of the age distribution of low mass stars in the solar
neighbourhood. 

Despite the large variety of SFRs there is a strong correlation to the AVR. 
For each SFR the slope and maximum velocity dispersion of the AVR are well
determined. 
For the AVR we find a power law with indices between 0.375 and 0.5.
The range of models is consistent with the
results of \citet{bin00} for the local age distribution and \citet{aum09} for
the SFR.

Applying the stellar lifetimes and the new scale height corrections to the PDMF
results in an IMF that shows only one break point at $1.7\msun$ and a steep
falloff at high masses. 

The density profile of an isothermal thick disc component can be fitted very
good by a sech$^\alpha_\mathrm{t}$ profile, where $\alpha_\mathrm{t}$
depends on the velocity dispersion.
The most prominent effect
of thick disc stars is the enhancement of the high velocity wings of 
K and M dwarfs in the range of 40--60\,km/s. From that we can exclude a heavy
thick disc with distinct kinematics and more than 10\% contribution to the local
stellar density.
Changing the thick disc parameters leads to slight variations of the
thin disc properties (mainly by assigning part of the stellar disc to the thick
disc), but has a negligible influence on the normalised 
SFR and AVR of the thin disc.

A variety of predictions can be made from the new disc model. 
The density profiles of MS star sub-populations depend on the lifetime of the
stars and are significantly different to density
profiles of single age sub-populations. The shape is neither exponential nor
sech$^2$ and can be characterised by the (half-)thickness and the exponential
scale height. From the
vertical density profiles MS stars number densities as function of colour and
apparent magnitude can be predicted. Applying these number densities with
observed 'Hess' diagrams from large surveys like the catalogues of the Sloan
Digital Sky Survey SEGUE/SDSS
enables us to restrict the parameters of the SFR further.
We determined vertical gradients in the kinematics which will be tested with
Radial Velocity Experiment (RAVE) data.
Age and metallicity distributions of stellar sub-populations as a function of $z$
above the galactic plane are predicted.

The future plan is to extend the local model to a complete disc model of the
Milky Way that provides a fully self-consistent connection of stellar
densities and kinematics. Ultimately this kind of detailed model is essential
to understand the large data sets as already available from SDSS and which are
expected on a much higher level in amount and precision by PanSTARRS and the
astrometric Gaia satellite mission.

\section*{Acknowledgements}
 
We thank Andrea Borch for providing the stellar evolution data with the
PEGASE code.
This research was supported in part by the National Science Foundation
under Grant No. PHY05-51164 via the KITP-program 'Building the Milky Way'
in Santa Barbara.


\label{lastpage}


\begin{thebibliography}{}
\bibitem[Abazajian et al.(2009)]{aba09} Abazajian, K. N., 
Adelman-McCarthy, J. K., Ag\"ueros, M. A. K., et al.\
	2009, ApJS, 182, 543
\bibitem[Andersen (1991)]{and91}
Andersen, J., 1991, A\&ARev. 3, 91
\bibitem[Aumer \& Binney (2009)]{aum09}
Aumer M., Binney J.J. 2009, MNRAS, 397, 1286
\bibitem[Bahcall(1984a)]{bah84a}
Bahcall, J.~N. 1984a, ApJ, 276, 156 
\bibitem[Bahcall(1984b)]{bah84b}
Bahcall, J.~N. 1984b, ApJ, 276, 169 
\bibitem[Bahcall \& Soneira(1980a)]{bah80a}
Bahcall, J.~N., Soneira, R.~M. 1980, ApJS, 44, 73 
\bibitem[Bahcall \& Soneira(1980b)]{bah80b}
Bahcall, J.~N., Soneira, R.~M. 1980, ApJ, 238, L17 
\bibitem[Bahcall \& Soneira(1984)]{bah84c}
Bahcall, J.~N., Soneira, R.~M. 1984, ApJS, 55, 67 
\bibitem[Bailer-Jones (2005)]{bai05}
Bailer-Jones, C. A. L., 2005,
in 'Transits of Venus: New Views of the Solar System and Galaxy', 
Proc. of IAU Coll. 196, ed. D.W. Kurtz,
Cambridge University Press, 429
\bibitem[Bensby et al.(2007)]{ben07}
Bensby, T., Zenn, A. R., Oey, M. S., Feltzing, S. 2007,
ApJ 663, L13
\bibitem[Bertelli et al.(1994)]{ber94}
Bertelli, G., Bressan, A., Chiosi, C. et al. 1994, A\&AS, 106, 275
\bibitem[Binney et al.(2000)]{bin00}
Binney, J., Dehnen, W., Bertelli, G. 2000,
MNRAS 318, 658
\bibitem[Bronfman et al.(1988)]{bro88}
Bronfman L., Cohen R.~S.,  Alvarez H., May J., Thaddeus P. 1988,
ApJ 324, 248
\bibitem[Cignoni et al.(2006)]{cig06}
Cignoni M., Degl'Innocenti S., Prada Moroni P.~G., Shore S.~N. 2006, 
A\&A 459, 783
\bibitem[Dehnen \& Binney(1998)]{deh98}
Dehnen, W., Binney, J. 1998, MNRAS 298, 387
\bibitem[Delhaye(1965)]{del65}
Delhaye, J. 1965, in Galactic Structure, Stars and Stellar Systems 5, 61
\bibitem[R. \& C. de la Fuente Marcos(2004)]{fue04}
de la Fuente Marcos R., de la Fuente Marcos C. 2004, NewA 9, 475 
\bibitem[Dickey \& Lockman(1990)]{dic90}
Dickey J.~M., Lockman F.~J. 1990,
ARAA 28, 215
\bibitem[Eggleton et al.(1989)]{egg89}
Eggleton P.~P., Fitchett M.~J., Tout C.~A. 1989, ApJ, 347, 998
\bibitem[Fioc \& Rocca-Volmerange(1997)]{pegase}
Fioc M., Rocca-Volmerange B. 1997, A\&A 326, 950
\bibitem[Freeman(1991)]{fre91}
Freeman, K.~C. 1991, in Sundelius B., Dynamics of Disc Galaxies. G\"oteborg
Observatory, G\"oteborg, 15
\bibitem[Girardi et al.(2002)]{gir02}
Girardi L., Bertelli G., Bressan A., Chiosi C., Groenewegen M.A.T., Marigo P.,
 Salasnich B., Weiss A. 2002, A\&A 391, 195
\bibitem[Girardi et al.(2004)]{gir04}
Girardi L., Grebel E.~K., Odenkirchen M., Chiosi C. 2004,
A\&A 422, 205
\bibitem[Haywood  et al.(1997a)]{hay97a}
Haywood M., Robin A. C., Cr\'eze M. 1997a,
A\&A 320, 428
\bibitem[Haywood  et al.(1997b)]{hay97b}
Haywood M., Robin A. C., Cr\'eze M. 1997b,
A\&A 320, 440
\bibitem[Haywood (2006)]{hay06}
Haywood M. 2006,
MNRAS 371, 1760
\bibitem[Henry \& Mc Carthy(1993)]{hen93}
Henry T.~J., Mc Carthy D.~W. Jr. 1993, 
AJ 106, 773
\bibitem[Henry et al.(1999)]{hen99}
Henry T.~J., Franz O.~G., Wasserman L.~H.~L. et al. 1999,
ApJ 512, 864
\bibitem[Hernandez et al.(2000)]{her00}
Hernandez X., Valls-Gabaud D., Gilmore G. 2000,
MNRAS, 316, 605
\bibitem[Holmberg \& Flynn(2000)]{hol00}
Holmberg J., Flynn C. 2000,
MNRAS, 313, 209
\bibitem[Holmberg \& Flynn(2004)]{hol04}
Holmberg J., Flynn C. 2004,
MNRAS, 352, 440
\bibitem[Holmberg et al.(2009)]{hol09}
Holmberg J., Nordstr{\"o}m B., Andersen J., 2009,
A\&A, in press, arXiv:0811.3982 {\bf GCS2}
\bibitem[Humphreys \& Larsen(1995)]{hum95}
Humphreys R.~M., Larsen J.~A. 1995,
AJ 110, 2183
\bibitem[Jahrei{\ss} \& Wielen(1997)]{jah97}
Jahrei{\ss} H., Wielen R. 1997,
In B. Battrick, M. A. C. Perryman, eds., Proc. ESA SP-402
(Nordwijk, ESA), 675
\bibitem[Just \& Jahrei{\ss}(2009)]{jus09}
Just A., Jahrei{\ss} H. 2009,
in preparation ({\bf Paper II})
\bibitem[Just et al.(1996)]{jus96}
Just A., Fuchs B., Wielen R. 1996,
A\&A 309, 715                         
\bibitem[Just et al.(2006)]{jus06}
Just A., Moellenhoff C., Borch A. 2006,
A\&A 459, 703
\bibitem[Kroupa et al.(1993)]{kro93}
Kroupa P., Tout C.~A., Gilmore G. 1993,
MNRAS 262, 545
\bibitem[Kuijken \& Gilmore(1991)]{kui91}
Kuijken K., Gilmore G. 1991,
ApJ 367, L9
\bibitem[Lucy(2000)]{luc00}
Lucy L.B.2000,
MNRAS 318, 92
\bibitem[Nordstr{\"o}m et al.(2004)]{nor04}
Nordstr{\"o}m B., Mayor M., Andersen J., et al. 2004,
A\&A 418, 989, {\bf GCS1}
\bibitem[Perryman et al.(2001)]{per01}
Perryman, M. A. C., de Boer, K. S., Gilmore, G., H{\o}g, E., Lattanzi, M. G.,
 Lindegren, L., Luri, X., Mignard, F., Pace, O., de Zeeuw, P. T., 2001,
 A\&A 369, 339
\bibitem[Phleps et al.(2005)]{phl05}
Phleps S., Drepper S., Meisenheimer K., Fuchs B. 2005,
A\&A 443, 929 
\bibitem[Pont \& Eyer(2005)]{pon05}
Pont F., Eyer L.  2005
in Proc. Gaia Symposium 'The Three-Dimensional Universe with Gaia' (ESA SP-576),
Eds. C. Turon, K.S. O'Flaherty, M.A.C. Perryman, 187 
\bibitem[Press et al.(1992)]{pre92}
Press W. H., Teukolsky S. A., Vetterling W. T., Flannery B. P. (eds.),
Numerical Recipes, Cambridge University Press, 1992
\bibitem[Reddy et al.(1993)]{red03}
Reddy B.~E., Tomkin, J., Lambert D.~L., Allende Prieto C. 2003,
MNRAS 340, 304
\bibitem[Rocca-Pinto et al.(2000)]{roc00}
Rocca-Pinto H.~J., Scalo J., Maciel W.~J., Flynn C. 2000, 
A\&A 358, 869
\bibitem[Robin et al.(2003)]{rob03}
Robin A.~C., Reyl\'e C., Derri\`ere S., Picaud S. 2003,
A\&A 409, 523                         
\bibitem[Roskar et al.(2008)]{ros08}
Roskar R., Debattista V. P., Quinn T. R., Stinson G. S., Wadsley J. 2008,
ApJ 684, L79
\bibitem[Scalo(1986)]{sca86}
Scalo J.~M. 1986,                                       
Fundamentals of Cosmic Physics 11, 1
\bibitem[Schmidt-Kaler(1982)]{lan82}
Schmidt-Kaler T. 1982,
in Landolt-B\"ornstein Vol. 2, 4.1                     
\bibitem[Sch\"onrich \& Binney(2009)]{sch09}
Sch\"onrich R., Binney J. 2009,
MNRAS 396, 203
\bibitem[Sellwood \& Binney(2002)]{sel02}
Sellwood J. A., Binney J. 2002,
MNRAS 336, 785
\bibitem[van Leeuwen(2007)]{vle07}
van Leeuwen F. 2007, Hipparcos, the new Reduction of the Raw Data, Springer
Dortrecht
\bibitem[Vergely et al.(2002)]{ver02}
Vergely, J.-L., K\"oppen, J., Egret, D., Bienaym\'{e}, O. 2002, A\&A  390, 917
\bibitem[Vyssotsky(1963)]{vys63}
Vyssotsky A.~N. 1963,                                   
in Basic Astronomical Data, Stars and Stellar Systems 3, 192
\bibitem[Wilson \& Woolley(1970)]{wil70}
Wilson O., Woolley R. 1970,                                      
MNRAS 148, 463  
\bibitem[Wielen(1977)]{wie77}
Wielen R. 1977,                                         
A\&A 60, 263  
\bibitem[Wielen et al.(1996)]{wie96}
Wielen R., Fuchs B., Dettbarn C. 1996,                                         
A\&A 314, 438  
\bibitem[Zwitter et al. (2008)]{zwi08} Zwitter, T., Siebert, A., 
Munari, U., et al. 2008, 
AJ 136, 421


\end{thebibliography}
\end{document}